\documentclass[sigconf]{acmart}

\usepackage{listings}
\usepackage{color}
\usepackage{pifont}% http://ctan.org/pkg/pifont
\newcommand{\cmark}{\ding{51}}%
\newcommand{\xmark}{\ding{55}}%

\usepackage{adjustbox}

\definecolor{light-gray}{gray}{0.80}

\definecolor{codegreen}{rgb}{0,0.6,0}
\definecolor{codegray}{rgb}{0.5,0.5,0.5}
\definecolor{codepurple}{rgb}{0.58,0,0.82}
\definecolor{codered}{rgb}{1.0,0,0}
\definecolor{codeyellow}{rgb}{0.95,0.95,0.8}
\definecolor{backcolour}{rgb}{0.95,0.95,0.92}
 
\lstdefinestyle{mystyle}{
    backgroundcolor=\color{backcolour},   
    commentstyle=\color{codegreen},
    keywordstyle=\color{magenta},
    numberstyle=\tiny\color{codegray},
    stringstyle=\color{codepurple},
    basicstyle=\scriptsize,
    breakatwhitespace=false,         
    breaklines=true,
    breakindent=0pt,
    columns=flexible,
    captionpos=t,                    
    keepspaces=true,                 
    numbers=none,                    
    numbersep=5pt,                  
    showspaces=false,                
    showstringspaces=false,
    showtabs=false,                  
    tabsize=1
}

\usepackage{xcolor}

\definecolor{mGreen}{rgb}{0,0.6,0}
\definecolor{mGray}{rgb}{0.5,0.5,0.5}
\definecolor{mPurple}{rgb}{0.58,0,0.82}
\definecolor{backgroundColour}{rgb}{0.95,0.95,0.92}

\lstdefinestyle{CStyle}{
    backgroundcolor=\color{backgroundColour},   
    commentstyle=\color{mGreen},
    keywordstyle=\color{magenta},
    numberstyle=\tiny\color{mGray},
    stringstyle=\color{mPurple},
    basicstyle=\footnotesize,
    breakatwhitespace=false,         
    breaklines=true,                 
    captionpos=t,                    
    keepspaces=true,                 
    numbers=left,                    
    numbersep=5pt,                  
    showspaces=false,                
    showstringspaces=false,
    showtabs=false,                  
    tabsize=2,
    language=C
}

\lstdefinelanguage
   [x64]{Assembler}     % add a "x64" dialect of Assembler
   [x86masm]{Assembler} % based on the "x86masm" dialect
   {morekeywords={jmpq, qword, push, pushq, jmp, CDQE,CQO,CMPSQ,CMPXCHG16B,JRCXZ,LODSQ,MOVSXD, %
                  POPFQ,PUSHFQ,SCASQ,STOSQ,IRETQ,RDTSCP,SWAPGS, %
                  rax,rdx,rcx,rbx,rsi,rdi,rsp,rbp, rip %
                  r8,r8d,r8w,r8b,r9,r9d,r9w,r9b, %
                  r10,r10d,r10w,r10b,r11,r11d,r11w,r11b, %
                  r12,r12d,r12w,r12b,r13,r13d,r13w,r13b, %
                  r14,r14d,r14w,r14b,r15,r15d,r15w,r15b}} % etc.
\usepackage{tikz}
\newcommand*\circled[1]{\tikz[baseline=(char.base)]{
            \node[shape=circle,draw,inner sep=0.2pt] (char) {#1};}}
\newcommand{\tikzcircle}[2][gray,fill=gray]{\tikz[baseline=-0.5ex]\draw[#1,radius=#2] (0,0) circle ;}%
\usepackage{boldline, multirow}
\usepackage[normalem]{ulem}
\useunder{\uline}{\ul}{}
\usepackage{subcaption}

\usepackage{cleveref}

\usepackage{dblfloatfix}

\usepackage{tabularx}

\newcommand\clearrow{\global\let\rowmac\relax}
\usepackage[utf8]{inputenc}
\usepackage{tikz,float}
\usepackage{tabularx}
\usepackage{xparse}

\NewDocumentCommand{\statcirc}{ O{#2} m }{%
    \begin{tikzpicture}
    \fill[#2] (0,0) circle (1.0ex); % Fill circle with base colour (arg#2)
    \fill[#1] (0,0) -- (90:1ex) arc (90:270:1ex) -- cycle; % Fill a half circle filled with second colour (arg#1), if specified
    \end{tikzpicture}
}

\usepackage{slashbox}
\usepackage{multicol}
\usepackage{amsmath}
\usepackage{algorithmic}
\usepackage{colortbl}
\usepackage[first=0,last=9]{lcg}
\definecolor{new-light-gray}{gray}{0.95}
\newcolumntype{g}{>{\columncolor{new-light-gray}}c}

\newtheorem{mydef}{\normalfont {\bf Definition}}
\usepackage{cramped}

\AtBeginDocument{%
  \providecommand\BibTeX{{%
    \normalfont B\kern-0.5em{\scshape i\kern-0.25em b}\kern-0.8em\TeX}}}

\setcopyright{none}
\renewcommand\footnotetextcopyrightpermission[1]{}

\begin{document}

\title{Methodologies for Quantifying (Re-)randomization Security and Timing under JIT-ROP}


%CCSR2020: Uncomment for authors' affilications
%\iffalse

\author{Salman Ahmed$^*$, Ya Xiao$^*$, Kevin Snow$^\dagger$, Gang Tan$^\ddagger$, Fabian Monrose$^\mathsection$, Danfeng (Daphne) Yao$^*$}
%\author{Ya Xiao}
%\email{yax99@vt.edu}
%\author{Danfeng (Daphne) Yao}
%\email{danfeng@vt.edu}
\affiliation{%
  \institution{$^*$Computer Science, Virginia Tech, $^\dagger$Zeropoint Dynamics, LLC, $^\ddagger$Computer Science and Engineering, Penn State University, $^\mathsection$Computer Science, UNC at Chapel Hill}
  %\streetaddress{P.O. Box 1212}
}
\email{{ahmedms, yax99, danfeng}@vt.edu, kevin@zeropointdynamics.com, gtan@cse.psu.edu, fabian@cs.unc.edu}

\begin{abstract}
%-------------------------------------------------------------------------------
Just-in-time return-oriented programming (JIT-ROP) allows one to dynamically discover instruction pages and launch code reuse attacks, effectively bypassing most fine-grained address space layout randomization (ASLR) protection. 
%Despite existing demonstrations showing the specific scenarios and conditions for a feasible attack, it is still unclear to what extent fine-grained code randomization and re-randomization impact code reuse attacks from a quantitative measurement perspective. 
%In-depth questions have not been studied, such as {\em How do starting pointers in JIT-ROP impact gadget availability?}; {\em How would one compute the re-randomization interval effectively to defeat JIT-ROP attacks?} {\em What impact do fine-grained randomization and re-randomization have on the Turing completeness of JIT-ROP payloads?}  
%Performing such a measurement under varying conditions requires a scalable approach to reproduce JIT-ROP operations. 
However, in-depth questions regarding the impact of code (re-)randomization on code reuse attacks have not been studied. For example, 
{\em how would one compute the re-randomization interval effectively by considering the speed of gadget convergence to defeat JIT-ROP attacks?}; 
{\em how do starting pointers in JIT-ROP impact gadget availability and gadget convergence time?};
%{\em what impact do re-randomization intervals have on the convergence of gadgets toward being Turing-complete?}; 
%{\em what impact do code locations have on the speed of convergence?}; 
{\em what impact do fine-grained code randomizations have on the Turing-complete expressive power of JIT-ROP payloads?}  
We conduct a comprehensive measurement study on the effectiveness of fine-grained code randomization schemes, with 
%a range of fine-grained randomization and re-randomization 
5 tools, 20 applications including 6 browsers, 1 browser engine, and 25 dynamic libraries. We provide methodologies to measure JIT-ROP gadget availability, quality, and their Turing-complete expressiveness, as well as to empirically determine the upper bound of re-randomization intervals in re-randomization schemes \color{black}using the Turing-complete (TC), priority, MOV TC, and payload gadget sets. Experiments show that the upper bound ranges from 1.5 to 3.5 seconds in our tested applications. Besides, our results show that locations of leaked pointers used in JIT-ROP attacks have no impacts on gadget availability, but have an impact on how fast attackers find gadgets\color{black}. Our results also show that instruction-level single-round randomization thwarts current gadget finding techniques under the JIT-ROP threat model.

\end{abstract}

\maketitle

% for removing running header
\pagestyle{plain}

%-------------------------------------------------------------------------------
\section{Introduction}
%-------------------------------------------------------------------------------
Just-in-time return-oriented programming (JIT-ROP) (e.g., ~\cite{snow2013just}) is a powerful attack technique that enables one to reuse code even under fine-grained address space layout randomization (ASLR). Fine-grained ASLR, also known as fine-grained code diversification or randomization, reorders and relocates program elements. Fine-grained randomization would defeat conventional ROP code reuse attacks~\cite{shacham2007geometry}, as the attacker no longer has direct access to the code pages of the victim program and its libraries. In other words, a leaked pointer only unlocks a small portion of the code region under fine-grained code randomization, seriously limiting the attack's ability to harvest code for ROP gadget purposes.

JIT-ROP attacks have the ability to discover new code pages dynamically~\cite{snow2013just}, by leveraging control-flow transfer instructions, such as {\em call} and {\em jmp}. Under fine-grained code randomization, the execution of a JIT-ROP attack is complex, as code page discovery has to be performed at runtime. From the defense perspective, re-randomization techniques (TASR~\cite{bigelow2015timely}, Shuffler~\cite{williams2016shuffler}, Remix~\cite{chen2016remix}, CodeArmor~\cite{chen2017codearmor}, RuntimeASLR~\cite{lu2016make}, and Stabilizer~\cite{curtsinger2013stabilizer}) have the potential to defeat JIT-ROP attacks. Besides, protections related to memory permission such as XnR~\cite{backes2014you}, NEAR~\cite{werner2016no}, Readactor~\cite{crane2015readactor}, destructive read such as Heisenbyte~\cite{tang2015heisenbyte}, and pointer indirection such as Oxymoron~\cite{backes2014oxymoron} specifically aim to thwart JIT-ROP attacks. Precise implementation of Control-Flow Integrity (CFI) can protect an application from all control-oriented attacks. The recently proposed Multi-Layer Type Analysis (MLTA)~\cite{lu2019does} technique improves CFI precision greatly by improving the accuracy in identifying indirect call targets. %A precise implementation of CFI is greatly fostered by the recently proposed technique called Multi-Layer Type Analysis (MLTA)~\cite{lu2019does} to identify indirect call targets. 

Even though the great promise of CFI for protecting control-oriented attacks, attackers may find ways to launch new exploits such as control-oriented~\cite{farkhani2018effectiveness, conti2015losing, goktas2014out,schuster2015counterfeit} and non-control-oriented~\cite{carlini2015control, hu2016data, ispoglou2018block} exploits as demonstrated before, where the exploits conform with CFI. A prime requirement of many of these exploits is information/pointer leakage. Thus, a measurement mechanism to design risk heuristics-based pointer selection and prioritization techniques is necessary for protecting pointers from leakage. Besides, from a {\bf defense-in-depth} perspective, it is important for a critical system to deploy multiple complementary security defenses in practice. A single defense may fail due to deployment issues such as implementation flaws or configuration issues. Thus, despite the strong security guarantees of CFI, our ASLR investigation is still extremely necessary.

%Besides, our re-randomization study can complement CFI.

Re-randomization techniques continuously shuffle the address space at runtime. This continuous shuffling breaks the runtime code discovery process by making the already discovered code pages obsolete. However, the interval between two consecutive randomizations must satisfy both performance and security guarantees.

Quantitative evaluation of how code (re-)randomization impacts code reuse attacks, e.g., in terms of interval choices, gadget availability, gadget convergence, and speed of convergence has not been reported. 
\color{black}
We define {\em gadget convergence} as the attack stage where an attacker has collected all the necessary gadgets. For example, if an attacker has found at least one gadget for each type of Turing-complete (TC) operations, then the gadget set is TC convergence. TC operations include memory, assignment, arithmetic, logic, control flow, function call, and system call~\cite{roemer2012return}.
%We use the term {\em gadget convergence} for a set of gadgets to indicate that the set of gadgets includes all the gadget types from a particular gadget set. 
%For example, a set of gadgets is converged to the Turing-complete gadget set if the set of gadgets includes all gadget types from the Turing-complete gadget set to cover the Turing-complete operations such as memory, assignment, arithmetic, logic, control flow, function call, and system call~\cite{roemer2012return}.
\color{black} 

(Re-)randomization techniques make it difficult for current gadget finding techniques to discover all gadgets. Thus, in-depth and systematic measurement is necessary, which can provide new insights on the impact of code (re-)randomization on various attack elements, such as code pointer leakage, various gadget sets, and gadget chain formation. It is also important to investigate how to systematically compute an effective re-randomization interval. Current re-randomization literature does not provide a concrete methodology for experimentally determining an upper bound of re-randomization intervals. Shorter intervals (e.g., millisecond-level) incur runtime overhead whereas longer intervals (e.g., second-level) give attackers more time to launch exploits. An upper bound would help guide defenders to make informed interval choices.

%It is also important to investigate how to systematically compute an effective re-randomization interval. Current re-randomization literature does not provide a concrete methodology for experimentally determining an upper bound of re-randomization intervals. An upper bound of re-randomization intervals must hamper the convergence of gadgets of being Turing-complete. Shorter intervals (e.g., millisecond-level) have the potential to hamper the convergence but incur performance or runtime overhead whereas longer intervals (e.g., second-level) give attackers more time to launch exploits. An upper bound would help guide defenders to make informed re-randomization interval choices.
%
%Besides the impact of code (re-)randomization on the availability of gadgets, code transformation techniques may limit the availability of gadgets. This is why we also investigate the impact of code transformations such compiler optimization techniques on the availability of gadgets. 
%
%
%Such a systematic and broad evaluation would complement existing attack demonstrations, as the latter only aims at demonstrating specific scenarios and conditions where fine-grained code randomization fails. Measuring the effectiveness of ASLR in a broader context, beyond a concrete attack, would help one design more effective solutions. Such a measurement effort has not been reported in the ROP literature.
%
We report our experimental findings on re-randomization interval choices considering the speed of gadget convergence, code pointer leakage, gadget availability, and gadget chain formation, under fine-grained ASLR and re-randomization schemes. 

\color{black}
Launching exploits is not a feasible measurement methodology to evaluate ASLR's effectiveness, due to {\em i)} low scalability -- exploit payload is not platform or application portable, {\em ii)} failure to exploit does not necessarily mean security, and {\em iii)} low reproducibility\color{black}. Our evaluation involves up to 20 applications, including 6 browsers, 1 browser engine, and 25 dynamic libraries.

We designed a measurement mechanism that allows us to perform JIT-ROP's code page discovery in a scalable fashion. This mechanism enables us to compare results from a number of programs and libraries under multiple ASLR conditions (coarse-grained, fine-grained function level, fine-grained basic block level, fine-grained instruction, and register levels). 
Our key experimental findings and technical contributions are summarized as follows.

\begin{itemize}
%Daphne commented below out, due to space. may include in the journal version.
%\item \textbf{\textit{A multi-step attack workflow that captures the common tasks and goals in ASLR bypasses}}. 

%\item \textbf{\textit{New definitions, metrics, and measurement methodologies.}} We define multiple new concepts, e.g., minimum footprint gadgets, extended footprint gadgets, and quality of gadgets, and describe methods for evaluating important properties of ROP gadgets, e.g., register corruption rate. We also summarize and experiment with common and specialized gadget types used in recent attacks. These contributions are useful beyond this specific ASLR study. 

\item 
We provide a methodology to compute the upper bound $\mathcal{T}$ for re-randomization intervals. \color{black}
%The upper bound $\mathcal{T}$ ensures that 
If the re-randomization interval is less than $\mathcal{T}$, then a JIT-ROP attacker is unable to obtain various gadget sets such as the Turing complete gadget set, priority gadget set, MOV TC gadget set, and gadgets from real-world payloads (see the definitions of gadget sets in Section~\ref{threat-model-assumptions}). We compute the upper bound T by measuring the minimum
time for an attacker to find a specific gadget set, i.e., the shortest time
to reach gadget convergence for the gadget set. The upper bound ranges from 1.5 to 3.5 seconds in our tested applications such as {\em nginx}, {\em proftpd}, {\em firefox}, etc.
\color{black}

\item 
%Locations of pointer leaks do not have an impact on the availability of gadgets. 
Our findings show that starting code pointers do not have any impact (i.e., zero standard deviations) on the reachability from one code page to another. Every code pointer leak is equally viable for derandomizing an address space layout, suggesting that an attacker's discovered gadgets eventually converge to a gadget set no matter where the starting pointer is. 

\item
Our findings also show that the starting code pointers have an impact on the speed of convergence. The time for a JIT-ROP attacker to discover a gadget set varies with the locations of starting code pointers. \color{black}In our experiments, the time for obtaining the Turing-complete gadget set ranges from 2.2 and 5.8 seconds.\color{black}

%CCSR2020 Our findings also show that the starting code pointers have an impact on the speed of convergence. That means the time for a JIT-ROP attacker to discover a gadget set varies. \color{blue}In our experiments, the fastest and average times for obtaining the Turing-complete gadget set are 2.2 and 4.3 seconds, respectively, for priority gadget set 1.5 and 3.5 seconds, for MOV TC gadget set 3.5 and 5.3 seconds, and for real-world payloads 3.3 and 4.9 seconds.\color{black}

%A pointer leakage in any location allows attackers to obtain a basic set of gadgets.

\item
We also present a general methodology for quantifying the number of JIT-ROP gadgets. Our results show that a single-round instruction-level randomization scheme can limit the availability of gadgets up to 90\% and break the Turing-complete operations of JIT-ROP payloads. Also, fine-grained randomization slightly degrades the gadget quality, in terms of register-level corruption. A stack has a higher risk of revealing dynamic libraries than a heap or data segment because our experiments show that stacks contain 16 more {\em libc} pointers than heaps or data segments on average. %This finding indicates the necessity of randomizing stack over heap or global variables.

\end{itemize}

Besides, %the comprehensive measurement work, 
we distill common attack operations in existing ASLR-bypassing ROP attacks (e.g.,~\cite{snow2013just, carlini2014rop, bittau2014hacking, davi2015isomeron}) and present a generalized attack workflow that captures the tasks and goals. This workflow is useful beyond the specific measurement study.

\section{Threat Model and Definitions}
\label{threat-model-assumptions}

%Daphne commented below out, as the attack tree is in the appendix
%Our threat model is illustrated by the two attack conditions $AC_1$ and $AC_3$ (highlighted in red) in our attack tree in Figure \ref{attac-decsision}. The most important defense in the two highlighted attack conditions ($AC_1$ and $AC_3$) is ASLR \textemdash both coarse- and fine-grained. Other standard defenses in the paths include W$\oplus$X, PIE, and RELRO, which are briefly described next.\footnote{A brief overview of these defenses is as follows (details in \S\ref{security-measures} in the appendix).} 

Coarse-grained ASLR (or traditionally known as only ASLR~\cite{team2003pax}) randomly relocates shared libraries, stack, and heap, but does not effectively relocate the main executable of a process.  This defense only ensures the relocation of the base address of a segment or module. The internal layout of a segment of the module remains unchanged. The \textbf{P}osition \textbf{I}ndependent \textbf{E}xecutable (PIE) option allows the main executable to be run as position independent code, i.e., PIE relocates the code and data segments. For comparison purposes, we performed experiments on coarse-grained ASLR with PIE enabled on a 64-bit Linux system.

Fine-grained ASLR, aka fine-grained code randomization or code diversification, attempts to relocate all the segments of the main executable of a process, including shared libraries, heap, stack, and memory-mapped regions and restructures the internal layouts of these segments. The granularity of the randomization varies, e.g., at the level of functions~\cite{conti2016selfrando,giuffrida2012enhanced, kil2006address}, basic blocks~\cite{chen2016remix, koo2018compiler, wartell2012binary}, instructions~\cite{hiser2012ilr}, or machine registers~\cite{homescu2013profile, crane2015readactor}. We evaluated randomization schemes at various levels of granularities using Zipr\footnote{https://git.zephyr-software.com/opensrc/irdb-cookbook-examples}~\cite{hawkins2017zipr}, Selfrando\footnote{https://github.com/immunant/selfrando} (SR)~\cite{conti2016selfrando}, Compiler-assisted Code Randomization\footnote{https://github.com/kevinkoo001/CCR} (CCR)~\cite{koo2018compiler}, and Multicompiler\footnote{https://github.com/securesystemslab/multicompiler} (MCR)~\cite{homescu2013profile}. We also evaluated Shuffler~\cite{williams2016shuffler}, a re-randomization tool. We are unable to test other tools due to various robustness and availability issues. %Listing \ref{list_sr_jmp_insertion} gives an example of the jump-based function repositioning technique used in Selfrando~\cite{conti2016selfrando}. 

We assume standard defenses such as W$\oplus$X and RELRO are enabled. W$\oplus$X specifies that no address is writable and executable at the same time. RELRO stands for Relocation Read Only. It ensures that the Global Offset Table (GOT) entries are read-only. RELRO is now by default deployed on mainstream Linux distributions.
%CCS20% In addition, our experimental evaluation is conducted under the following assumptions. Attackers do not have any prior knowledge of the target application's memory layout, i.e., attackers have to derandomize the layout through an attack. 
%CCS20% Fine-grained code randomization is applied in every executable and associated library in a target system (unless specified otherwise). 
%Commented out by Daphne
%We assume the target system is not equipped with control-flow integrity (CFI), as an ideal form of CFI \cite{abadi2005control} prevents most of the code-reuse attacks that subvert the control flow of a program. 
%

%\color{blue}
\noindent
\textbf{\textit{Layered defenses}.}
%ask to discuss their effectiveness and justify the omission of layered defenses such as CFI, CPI, and memory permission protections in our ASLR measurement.
%Response: 
CFI and Code Pointer Integrity (CPI) solutions are very powerful techniques. Yet, it is still necessary for one to experimentally measure the effectiveness of various defense implementations in practice (e.g., CPI enforcement with spatial and temporal guarantees, CFI implementations with various granularities like~\cite{burow2017control}). From a measurement perspective, it is useful and necessary to isolate various defense factors. Decoupling them helps one better understand the individual factor's security impact. Otherwise, it might be too complicated to interpret the experimental results. This is the reason we chose to focus on ASLR defenses in this work and omit other defenses (e.g, CFI~\cite{abadi2005control, zhang2013practical, zhang2013control, niu2014modular, criswell2014kcofi, payer2015fine, mohan2015opaque, ghaffarinia2019binary} and CPI~\cite{backes2014oxymoron, cowan2010pointguard, evans2015missing, kuznetsov2014code, kuznetzov2018code, lu2015aslr,  mashtizadeh2015ccfi}. For similar reasons, we also omit memory permission protections (e.g., XnR~\cite{backes2014you}, NEAR~\cite{werner2016no}, Readactor~\cite{crane2015readactor} and Heisenbyte~\cite{tang2015heisenbyte}) for this paper. Execute-only-Memory (XOM\footnote{XoM is now supported natively at the hardware level on x86 systems with memory protection keys (MPK) support and Armv7-M or Armv8-M processors.})~\cite{lie2000architectural} and Execute-no-Read (XnR)~\cite{backes2014you} style defenses are also powerful. But, attacks such as AOCR~\cite{rudd2017address} and code inference~\cite{snow2016return} are still possible with these defenses. %But these attacks are not JIT-ROP style, hence not related to the desired study of this paper. 
We also discuss the need for measuring code pointer protection solutions under the JIT-ROP model in Section~\ref{sec:discussion}. %In our ongoing work, we are conducting measurements on CFI and CPI defenses. 

We assume attackers have already obtained a leaked code pointer (e.g., a function or a virtual table pointer) through remote exploitation of a vulnerability. Such an assumption is standard in existing attack demonstrations. Also, fine-grained code randomization is applied in every executable and associated library in a target system (unless specified otherwise). \color{black}A JIT-ROP attacker knows nothing about the applied fine-grained randomization.\color{black}
%\color{black}

%\color{blue}
\noindent
{\textbf{Native vs. WebAsm vs. JavaScript version of JIT-ROP}.}
While the original JIT-ROP attack was demonstrated in a browser using JavaScript, the attack approach has general applicability in both native and scripting environments. Our experiments are focused on the native execution of JIT-ROP attacks. We conducted the experiments for measuring the re-randomization upper bound using the native JIT-ROP code module. The execution time of WebAssembly is within 10\% to 2x of native code execution~\cite{haas2017bringing}; JavaScript is on average 34\% slower than WebAssembly~\cite{haas2017bringing}. Thus, our re-randomization intervals measured using the native execution would be conservatively applicable for the scripting environments as well. \color{black}Besides, JIT-ROP is not related to the JIT compilers of JavaScript (JS) engines and does not use any flaws of JIT compilers to perform a code-reuse attack, though some work~\cite{athanasakis2015devil} uses such flaws. JIT-ROP harvests gadgets from a target binary's static code, which is finely randomized; it does not harvest gadgets from dynamically generated code (e.g., scripts). Thus, JS or WebAsm versions do not make substantial differences in gadget availability. \color{black}

%\color{black}

%Similar to JIT-ROP \cite{snow2013just}, we assume that no code pointer protection \cite{backes2014oxymoron, cowan2010pointguard, evans2015missing, kuznetzov2018code, kuznetsov2014code,  lu2015aslr,  mashtizadeh2015ccfi} exists in the target application. We discuss the need for measuring code pointer protection solutions under the JIT-ROP model in Section~\ref{sec:discussion}. We assume that memory permission-related protections such as XnR~\cite{backes2014you}, NEAR~\cite{werner2016no}, Readactor~\cite{crane2015readactor} and destructive read-related protections such as Heisenbyte~\cite{tang2015heisenbyte}, etc.
%, and re-randomization (TASR~\cite{bigelow2015timely}, Shuffler~\cite{williams2016shuffler}, Remix~\cite{chen2016remix}) related defenses 
%
%are not present in the victim machine\footnote{Attacks (e.g, AOCR~\cite{rudd2017address} and code inference \cite{snow2016return}) are still possible with those defenses.}. We assume attackers have already obtained a leaked code pointer (e.g., a function pointer or a virtual table pointer) through remote exploitation of an application/library vulnerability. Such an assumption is standard in existing attack demonstrations. 
%We discuss its difficulty in Section~\ref{eval-RQ3}. \gtan{What difficulty?}

%\color{blue}
Next, we discuss the terms of Turing-complete gadget set, priority gadget set, MOV TC gadget set, re-randomization upper bound, minimum footprint gadgets, and extended footprint gadgets.
%\color{black}
%{\bf Salman, can you make the 2 definitions italic? see some of my papers for an example}

\begin{mydef}\label{turing-completeness}
{\em Turing-complete gadget set refers to a set of gadgets that covers the Turing-complete operations including memory operations (i.e., load memory LM and store memory SM gadgets), assignments (i.e., load register LR and move register MR gadgets), arithmetic operations (i.e., arithmetic AM, arithmetic load AM-LD, and arithmetic store AM-ST gadgets), logical operations (i.e., logical gadgets), control flow (i.e., jump JMP gadgets), function calls (i.e., CALL gadgets), and system calls (i.e., system SYS gadgets)~\cite{roemer2012return}.}
\end{mydef}

\begin{mydef}\label{upperbound}
{\em The upper bound $\mathcal{T}_\mathcal{P}^\mathcal{A}$ of a re-randomization scheme $\mathcal{P}$ under a JIT-ROP attacker $\mathcal{A}$ is the maximum amount of time between two consecutive randomization rounds that prevents $\mathcal{A}$ from obtaining a Turing-complete, priority, MOV TC, or payload gadget set, i.e., for any interval $\mathcal{T'}_\mathcal{P}^\mathcal{A} < \mathcal{T}_\mathcal{P}^\mathcal{A}$,  the set of gadgets obtained under $\mathcal{T'}_\mathcal{P}^\mathcal{A}$ does not converge to any of the four gadget sets.}
\end{mydef}

\textbf{Extended footprint (EX-FP) gadgets}: 
%Turing-complete gadgets (i.e., load, store, assignment, etc.) and attack-specific gadgets (e.g., reflector gadget, call site gadget, etc.) are useful for arbitrary computation and building an attack payload. 
A gadget is an extended footprint gadget if it is an instance of the Turing-complete gadget set or an instance of attack-specific gadgets. An EX-FP gadget may contain additional instructions that may cause side effects in an attack payload. \color{black}EX-FP gadgets include longer memory addressing expressions.\color{black}

\textbf{Minimum footprint (MIN-FP) gadgets}: A minimum footprint gadget is an instance of the Turing-complete gadget set or attack-specific gadgets without causing any side effects.
%in an attack payload.

%\color{blue}

Our definition of the Turing-complete gadget set represents our best efforts, by no means the only way. For example, a pair of load (LM) and store (SM) gadgets may potentially replace a move (MR) gadget. However, they may not be directly equivalent due to possibly mismatching memory offsets of EX-FP load gadgets or the scarcity of MIN-FP load gadgets. Excluding load-n-store from the Turing-complete gadget set might underestimate attackers' capabilities, while including them might overestimate attackers' capabilities. We perform our measurements considering the Turing-complete gadget set that enables the highest expressiveness of ROP attacks. %Besides, the Turing-complete gadget set would be highly desirable in the context of a JIT-ROP attack because of the nature of sophisticated payload generated at the time of the attack on the fly. 
However, under this condition, our results might underestimate the attackers' capabilities. \color{black}To balance an attacker's capabilities, we further break down the Turing-complete gadget set into two smaller gadget sets: {\em i)} priority gadget set and {\em ii)} MOV TC gadget set. The {\em priority} gadget set includes 10 most frequently used gadgets in 15 real-world ROP chains from Metasploit. The {\em MOV Turing-complete} gadget set~\cite{dolan2013mov} requires six MOV gadgets and four unique registers. Besides, we also include three real-world ROP payloads from Metasploit in our measurement. 
\color{black}

%CCSR202 To balance an attacker's capabilities, we further break down the Turing-complete gadget set into two smaller gadget sets and perform our measurements using these two sets. The two smaller gadget sets are i) priority gadget set and MOV Turing-complete gadget set. The priority gadget set includes 10 most frequently used gadgets in real-world ROP chains from Metasploit. The MOV Turing-complete gadget set~\cite{dolan2013mov} requires six MOV gadgets and four unique registers. Besides, both gadget sets require a system-call gadget.

New metrics proposed by Brown and Pande's~\cite{brown2019less} work -- functional gadget set expressivity and special-purpose gadget availability -- are new leads that will help relax the expressiveness condition of the Turing-complete gadget set in the future.

%CCS20% Our definition of Turing-complete gadget set represents our best efforts, by no means the only way. For example, a pair of load (LM) and store (SM) gadgets may potentially replace a move (MR) gadget. However, they may not be directly equivalent due to possibly mismatching memory offsets of EX-FP load gadgets or the scarcity MIN-FP load gadgets. Excluding load-n-store from the Turing-complete gadget set might underestimate attackers’ capabilities, while including them might overestimate attackers’ capabilities. Besides, the Turing-complete gadget set would be highly-desirable in the context of a JIT-ROP attack because of the nature of sophisticated payload generated at the time of the attack on the fly.
%\color{black}

Our security definition of the upper bound in Definition~\ref{upperbound} is specific to the JIT-ROP threat, and is not applicable to other threats (e.g., side-channel threats). A shorter interval may still allow attackers to gain information. However, as our Section~\ref{attack-workflow} shows, without gadgets that information may not be sufficient for launching exploits.
%------------------------------------------------------------------------------------
\section{JIT-ROP vs. Basic ROP Attacks}\label{attack-workflow}
%CCS20% \section{Comparison of JIT-ROP and Basic ROP Attacks}\label{attack-workflow}
%------------------------------------------------------------------------------------
\noindent We manually analyze a number of advanced attacks to extract common attack elements and identify unique requirements. We illustrate the key technical differences between JIT-ROP and conventional (or basic) ROP attacks. This section helps one understand our experimental design in Section~\ref{exp-design} and findings in Section~\ref{exp_eval}. We analyze various attack demonstrations with a focus on attacks (e.g.,~\cite{davi2015isomeron, snow2013just, bittau2014hacking, carlini2014rop}) in our threat model. 

%backes2014oxymoron, maisuradze2016cannot, carlini2015control, goktas2014out, hu2016data

%\onecolumn
\begin{figure}[!tbph]
  \centering
  %,natwidth=610,natheight=642
  \includegraphics[width=0.48\textwidth]{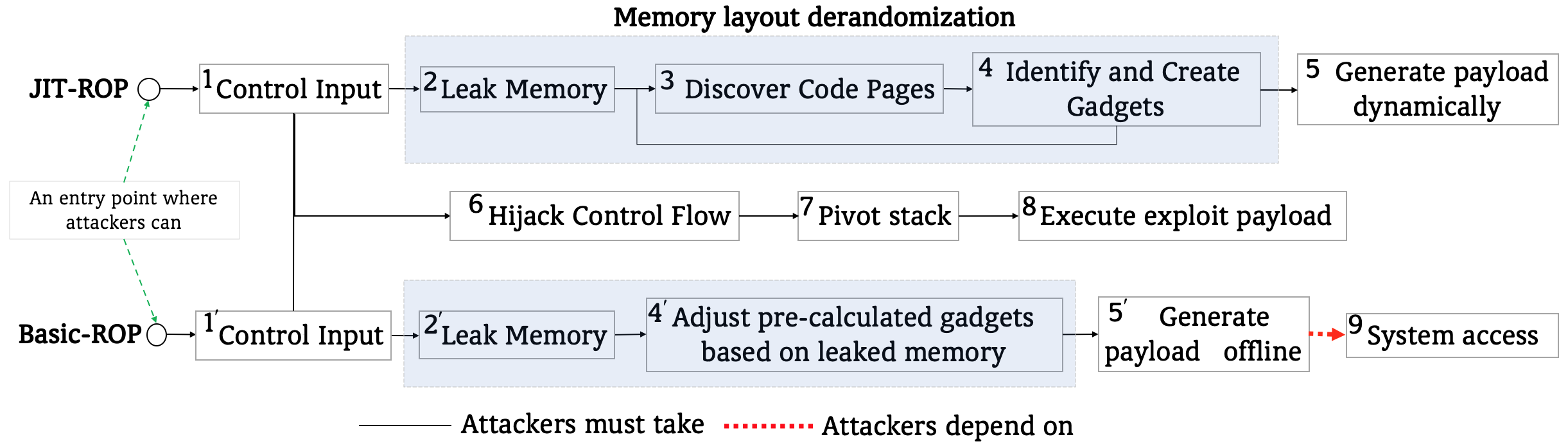}
  \caption{An illustration of the commonalities and differences between a conventional (or basic) ROP attack (bottom) and a JIT-ROP attack (top). The top gray-box highlights the key steps in JIT-ROP to overcome fine-grained ASLR.}
  \label{fig-attack-tree}
\end{figure}
%\twocolumn

To overcome both coarse- and fine-grained ASLR and conduct an attack to gain privileged operations, an attacker needs to perform the tasks presented in Figure~\ref{fig-attack-tree}.  The attack workflow has three major components: \textbf{\textit{memory layout derandomization}}, \textbf{\textit{system access}}, and \textbf{\textit{payload generation}}. %CCSR2020 We describe each component in Sections~\ref{workflow-memory-derand}, \ref{workflow-system-access}, and \ref{workflow-payload-gen}, respectively.

\subsection{Memory Layout Derandomization} \label{workflow-memory-derand}
%------------------------------------------------------------------------
\noindent Derandomizing an address space layout is the key for mounting code-reuse attacks. Due to the W$\oplus$X defense, attackers need to derandomize the memory layout to discover gadgets (steps \circled{2}-\circled{4} for JIT-ROP and steps \circled{2$'$} and \circled{4$'$} for basic ROP in Figure~\ref{fig-attack-tree}). Usually, attackers leverage memory corruption vulnerabilities to leak memory~\cite{strackx2009breaking} and derandomize an address space layout using the leaked memory. This step requires overcoming several obstacles.
%Usually, attackers leverage memory corruption vulnerabilities to leak memory \cite{strackx2009breaking} and start the derandomization process utilizing the leaked memory. This step requires overcoming several obstacles.

%-----------------------------------------------------------------------------------
\noindent
\textbf{Memory disclosure}.
\label{workflow-memory-derand-disclosure}
%-----------------------------------------------------------------------------------
The most common way of derandomizing memory layout is through a memory disclosure vulnerability. Attackers use vulnerabilities in an application's memory (e.g., heap overflows, use-after-free, type confusion, etc.) and weaknesses in system internals (e.g., vulnerabilities in the glibc malloc implementation or its variants~\cite{argyroudis2012exploiting, heelan2018automatic}, Heap Feng Shui~\cite{sotirov2007heap}, and Flip Feng Shui~\cite{razavi2016flip}) to leak memory contents (Steps \circled{2} and \circled{2$'$}). Details on memory corruption can be found in~\cite{szekeres2013sok, gens2018k} and an example in~\cite{strackx2009breaking}.

%------------------------------------------------------
\noindent
{\bf Code reuse}. \label{code-reuse}
%-----------------------------------------------------
Due to W$\oplus$X defense, adversaries cannot inject code in their payload. ROP~\cite{shacham2007geometry} and its variants Jump-Oriented Programming (JOP)~\cite{bletsch2011jump} and Call-Oriented Programming (COP)~\cite{goktas2014out} can defeat this defense. These techniques use short instruction sequences (i.e., gadget) from the code segments of a process' address space and allow an adversary to perform arbitrary computations.  ROP tutorials can be found in~\cite{snow2013just, davi2011ropdefender}. The difference between basic ROP~\cite{shacham2007geometry} and JIT-ROP~\cite{snow2013just} is described next.

\noindent
{\bf Basic ROP}. Coarse-grained ASLR only randomizes the base addresses of various segments and modules of a process. The content of the segments and modules remains unchanged. Thus, it is feasible for an adversary to launch a basic ROP attack~\cite{basic-rop} using gadgets given a leaked address from the code segment of interest. The adversary only needs to adjust the addresses of pre-computed gadgets w.r.t. the leaked address. Step \circled{4$'$} in Figure~\ref{fig-attack-tree} is about this task.

\noindent
{\bf Just-in-time ROP}. Fine-grained ASLR randomizes the base addresses, as well as the internal structures of various segments and modules of a process. Thus, simply adjusting the addresses of pre-computed gadgets (as in the basic ROP) no longer works. An attacker needs to find gadgets dynamically at the time of an exploit. Scanning a process' address space linearly for gadgets by starting from a disclosed code pointer may not be effective because this linear scanning may lead to 
%a segmentation fault and 
crash the process due to reading an unmapped memory.
%CCSR2020 She may attempt to scan the process address space starting from a disclosed code pointer to search for gadgets. However, this linear scanning may lead to a segmentation fault and crash the process due to reading an unmapped memory address. 
A powerful technique introduced in JIT-ROP~\cite{snow2013just} is recursive code page harvest, which is explained next.
%Adversaries also need techniques to identify the kind of a gadgets while searching for gadgets. This technique can also be applicable in the presence of coarse-grained randomization, but not necessary. 

The {\bf recursive code harvest} technique exploits the connectivity of code in memory to derandomize and locate instructions (step \circled{3} in Figure~\ref{fig-attack-tree}). The technique identifies gadgets at runtime by reading and disassembling the text segment of a process. The technique computes the page number from a disclosed code pointer and reads the entire 4K data of that page. A light-weight disassembler converts the page data into instructions. The code harvest technique searches for chain instructions, such as {\em call} or {\em jmp} instructions to find pointers to other code pages.

%CCSR2020 JIT-ROP attacks exploit the connectivity of code in memory to derandomize and locate instructions. The {\bf code harvest process} in JIT-ROP identifies gadgets at runtime by reading and disassembling the text segment of a process. This process starts with computing the page number of a disclosed code pointer at runtime. A 64-bit system uses the first 52 bits for page numbers if the page size is 4K. Once the page number is computed, the process reads the entire 4K data of that page. A light-weight disassembler converts the page data into instructions. The code harvest process searches for chain instructions, such as {\em call} or {\em jmp} instructions to find pointers to other code pages. 

\begin{figure}[!th]
  \centering
  \includegraphics[width=0.40\textwidth]{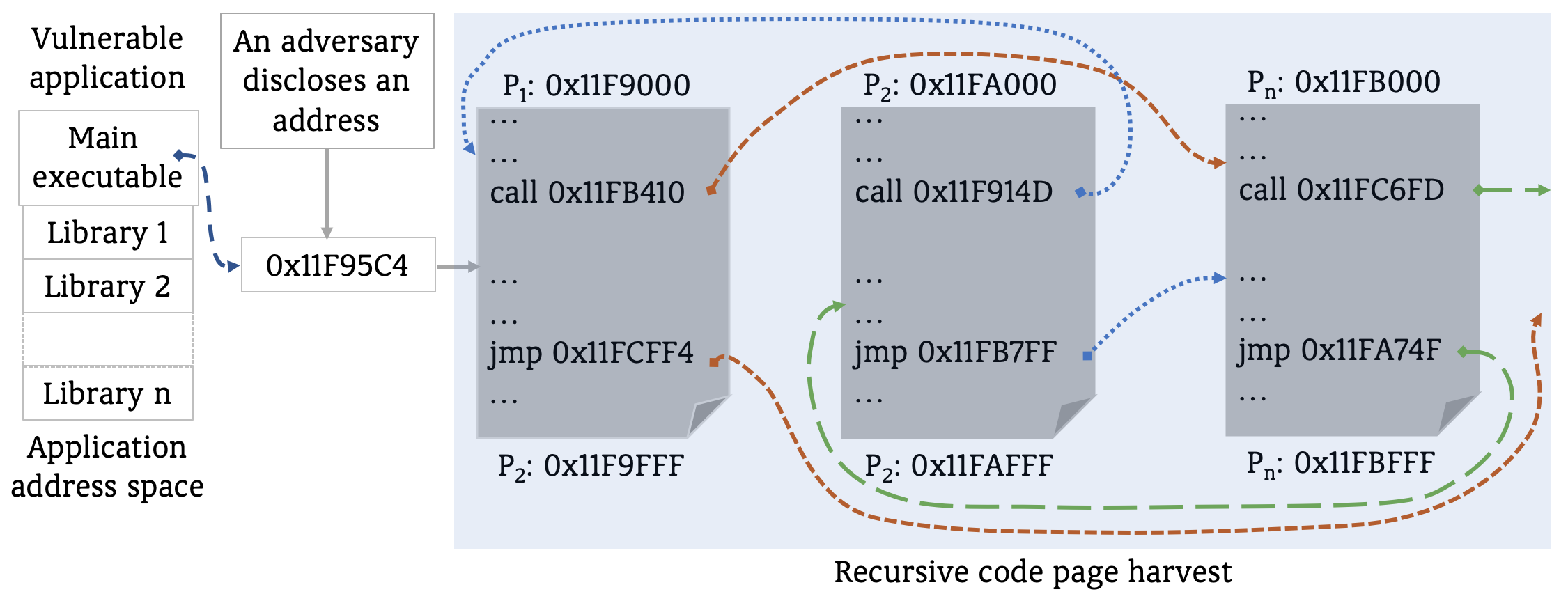}
  \caption{ An illustration of the recursive code harvest process of JIT-ROP~\cite{snow2013just}. An adversary discloses an address from the main executable or libraries (in this case from the main executable) of an application through a vulnerability.
}
  \label{cp-jitrop}
\end{figure}

An illustration is shown in Figure~\ref{cp-jitrop}. The code harvest process starts from the disclosed pointer (0x11F95C4), reads 4K page data (0x11F9000-0x11F9FFF), disassembles the data, searches for {\em call} and {\em jmp} instructions to find other pointers (0x11FB410 and 0x11FCFF4) to jump to those code pages. This process is recursive and stops when all the reachable code pages are discovered. %CCSR2020 We implemented this code harvesting method for our evaluation. It is important to mention that indirect calls to library functions can also be resolved to jump to the library code pages.

Snow {\it et al.} demonstrated the JIT-ROP attack in a browser. Since exploiting a memory corruption bug remotely covers a wide variety of exploits, a browser is an ideal interface for JIT-ROP attacks. 
The scripting environment of a browser enables easy interfacing of a JavaScript-based JIT-ROP attack payload. Similarly, JIT-ROP attack payload can be embedded into a PDF reader that supports JavaScript (e.g., Adobe Reader). However, %to launch a JIT-ROP attack in any scripting environment, 
an attacker must convert any non-scripting attack code to script for the scripting environment. For example, the original JIT-ROP framework was written in C/C++ and was transpiled to JavaScript to demonstrate on Internet Explorer.

%CCSR2020 Snow {\it et al.} demonstrated the JIT-ROP attack in a browser. Since exploiting a memory corruption bug remotely covers a wide variety of exploits, a browser could be an ideal interface for JIT-ROP attacks. More specifically, the scripting environment of a browser enables easy interfacing of a JIT-ROP attack payload. The scripting environment, for example, the JavaScript engine allows Just-In-Time compilation which converts JavaScript bytecode to native machine code with the help of the JavaScript virtual machine. That means the scripting environment enables executing dynamically generated code without any concerns related to the DEP or NX memory permission. This way a JIT-ROP JavaScript payload can be executed within a browser. Similarly, JIT-ROP attack payload can be embedded into a PDF reader that supports JavaScript (e.g, Adobe Reader). To launch a JIT-ROP attack in a scripting environment, an attacker must convert any non-scripting code for the scripting environment. For example, the original JIT-ROP framework was written in C/C++ and transpiled the framework to JavaScript to demonstrate an attack on a browser. A framework like Emscripten\footnote{https://emscripten.org} can transpiles C/C++ code to JavaScript/WebAssembly.

%\twocolumn

%----------------------------------------------------------------------------------------------------------
\noindent
%CCSR2020 \textbf{Code page harvest and gadget identification}. \label{workflow-memory-derand-harvest}
\textbf{Gadget identification}. \label{workflow-memory-derand-harvest}
%----------------------------------------------------------------------------------------------------------
%CCSR2020 In step \circled{3} of Figure \ref{fig-attack-tree}, an adversary utilizes the leaked memory address to read code pages of the process' code segments. In step \circled{4}, she can identify and create gadgets by disassembling the read code pages. She can scan for byte values corresponding to {\em ret} opcodes (e.g., 0xC2, 0xC3) and perform a narrow-scoped backward disassembly from there. %She also can use an improved disassembler (e.g., Capstone \cite{capstone-disas}) to perform gadget identification. 
In step \circled{4} of Figure~\ref{fig-attack-tree}, attackers identify gadgets by scanning for byte values corresponding to {\em ret} opcodes (e.g., 0xC2, 0xC3) from the read code pages and perform a narrow-scoped backward disassembly.
The adversary performs step \circled{3} and \circled{4} repeatedly to find required gadgets for the target exploit.

%------------------------------------------------------------
\subsection{System Access} \label{workflow-system-access}

\noindent Attackers need to issue system APIs or gadgets to perform privileged operations. If the CFI defenses %\footnote{which restrict "gadgets" to only legitimately called functions} 
(e.g., BCFT~\cite{ghaffarinia2019binary}, CCFIR~\cite{zhang2013practical} and bin-CFI~\cite{zhang2013control}) are not enforced, adversaries do not need to invoke the entire functions to ensure legitimate control flow. An adversary can just chain together enough gadgets for setting up the arguments of a system call and invoking it. This observation is particularly true for Linux, which is the focus of this paper. In Windows exploits~\cite{snow2013just}, the approach can be slightly different, as adversaries commonly invoke a system API instead of invoking a system call directly. {\em Syscall} gadgets can be found in an application's code or dynamic library. For basic ROP attacks, attackers can adjust pre-computed system gadgets from dynamic libraries, given that she manages to obtain a code pointer from a dynamic library (e.g., {\em libc}). Step \circled{9} in Figure~\ref{fig-attack-tree} is for this task. This task is performed manually and offline. The attacker may obtain the library code pointer from an application's stack or heap or data segment. One can find system gadgets through step \circled{4} in JIT-ROP.

\subsection{Payload Generation} \label{workflow-payload-gen}
%-----------------------------------------------------------
\noindent %CCS20%Once an adversary derandomizes the memory layout of a process and gets access to enough gadgets, she glues different parts (e.g., gadgets, functions, strings, constants, etc.) together to build a payload or attack chain. 
Attackers generate payloads by putting many pieces (e.g., gadgets, functions, constants, strings, etc.) together. This process must ensure a setup for calling system functions or system gadgets. An attacker generates a payload dynamically at step \circled{5} in the presence of fine-grained code randomization or manually at step \circled{5$'$} in the presence of coarse-grained code randomization and stores the payload in a stack/heap. Because a payload is primarily a set of addresses that point to some existing code in an application's address space, attacks do not execute anything stored in a stack/heap, which is protected by W$\oplus$X. An attacker may utilize the same vulnerability as in step \circled{2} or a different vulnerability to hijack a program's control flow at step \circled{6} to redirect the flow to the stored payload. The target of a payload is to achieve an attack goal, e.g., memory leak or launching a malicious application/root shell.
%CCSR2020% Once an adversary gets access to enough gadgets, she constructs a payload or gadget chain with an attack goal in mind. The adversary may generate the payload dynamically at step \circled{5} in the presence of fine-grained code randomization or manually at step \circled{5$'$} in the presence of coarse-grained code randomization and stores the payload in a stack. Because a payload is primarily a set of addresses that point to some existing code in an application's address space, attacks do not execute anything stored in a stack/heap, which is protected by W$\oplus$X. The adversary may utilize the same vulnerability as in step \circled{2} or a different vulnerability to hijack a program's control flow at step \circled{6} to redirect the flow to the stored payload.

It is desirable for attackers to obtain attack chains that have minimal side effects, i.e., having a payload that fulfills attack goals without generating any unnecessary computation. However, this property may not be guaranteed if the gadget availability is limited by code randomization. We refer to the side effect of gadgets as {\em footprints}.  We defined the {\em minimum footprint} gadget and {\em extended footprint} gadget in Section~\ref{threat-model-assumptions}.% and experimentally measure the quality of gadgets in \S\ref{eval-RQ5}.  

For ROP attacks (e.g.,~\cite{carlini2014rop}) that bypass control-flow integrity (CFI) defenses, the attackers also need to prepare specialized payloads in addition to the previous tasks. For example, the Flashing (FS) and Terminal (TM) gadgets in Table~\ref{gadget_type} in the Appendix were designed by Carlini and Wagner~\cite{carlini2014rop} to bypass specific CFI implementations (namely, kBouncer~\cite{pappas2013transparent} and ROPecker~\cite{cheng2014ropecker}).

%----------------------------------------------
%CCS20%\section{Experimental Design}\label{exp-design}
\section{Measurement Methodologies}\label{exp-design}
%----------------------------------------------
\noindent We describe our measurement methodologies for evaluating fine-grained ASLR's impact on the memory layout derandomization, system access, and payload generation of JIT-ROP. %We also explain the measurement challenges. We first define {\em extended footprint (EX-FP) gadgets}, {\em minimum footprint (MIN-FP) gadgets}, and the {\em register corruption rate} next.
One major challenge is how to {\bf quantify} the impact of fine-grained code randomization or re-randomization.  Our approach is to count the number of gadgets that are available to attackers under the JIT-ROP code harvest mechanism.
Other challenges are how to quantify {\em i)} the difficulty of accessing internal system functions and  {\em ii)} the quality of gadget chains.
For the former, our approach is to measure the number of system gadgets and count {\em libc} pointers in a stack or heap or data-segment of an application. In order to quantify the quality of gadget chains, we design a register-level measurement heuristic to compute the register corruption rate.
\subsection{Methodology for Derandomization} \label{exp-memory-derand}
%-----------------------------------------------------------------------------------------------
%CCS20%This methodology is for evaluating RQs \#1 and \#2. The main metric of this methodology is the number of various gadgets.
% (see the evaluation results in \S\ref{eval-RQ1}, \ref{eval-RQ4}, and \ref{eval-RQ5}, respectively).

\noindent
{\bf Gadget selection}. We manually extracted 21 types of gadgets from various attacks~\cite{snow2013just, carlini2015control, carlini2014rop, goktas2014out, bittau2014hacking}. These gadget types include load memory (\verb1LM1), store memory (\verb1SM1), load register (\verb1LR1), move register (\verb1MR1), arithmetic (\verb1AM1), arithmetic load (\verb1AM-LD1), arithmetic store (\verb1AM-ST1), \verb1LOGIC1, jump (\verb1JMP1), call (\verb1CALL1), system call (\verb1SYS1), and stack pivoting (\verb1SP1) gadgets. In addition to these, the gadget types also include some attack-specific gadgets such as call preceding (\verb1CP1), reflect (\verb1RF1), call site (\verb1CS21) and entry point (\verb1EP1) gadgets. Table~\ref{gadget_type} in the Appendix shows those gadget types in more details.
%Each row describes the gadget types, gadget's purpose, MIN-FP form of that gadget type, an example of that gadget type, and short name (SN) for our convenience. The gadget types include gadgets for both Turing-complete and attack-specific gadgets. 

These 21 types of gadgets include the Turing-complete gadget set (see Definition~\ref{turing-completeness}). \color{black}These gadgets also include the priority and MOV TC gadget sets (Table~\ref{tab:gadgets-priority-movtc-sets} in the Appendix). The Turing-complete, priority, and MOV TC gadget sets with some attack-specific gadgets (e.g., \verb1CP1, \verb1RF1, \verb1CS21, and \verb1EP1) are appropriate for our evaluation because we can precisely identify those gadgets\color{black}. Some attack specific gadgets such as \verb2CS12, \verb1FS1, etc. are very application-specific and do not have concrete forms or attack goals. These gadgets are used to trick defense mechanisms. We leave these gadgets out of our evaluation. \color{black}We also include gadgets from three real-world ROP payloads from Metasploit~\cite{payloadOne, payloadEight} and Exploit-Database~\cite{payloadFour}\color{black}. We discuss the evaluation of the block-oriented gadgets used for Block-Oriented Programming (BOP)~\cite{ispoglou2018block}.

%CCSR2020 Some attack specific gadgets such as \verb2CS12, \verb1FS1, etc. are very application-specific. These gadgets do not have any concrete forms or concrete attack goals. These gadgets are used to trick defense mechanisms. We leave these gadgets out of our evaluation. \color{blue}We also include gadgets from three real-world ROP payloads from Metasploit~\cite{payloadOne, payloadEight} and Exploit-Database~\cite{payloadFour}\color{black}. We also discuss the evaluation of the block-oriented gadgets used for Block-Oriented Programming (BOP)~\cite{ispoglou2018block}.

% (https://tinyurl.com/vdxhcab)

%Daphne commented out below
%The availability of the set of gadgets described in the table (Table \ref{gadget_type}) can be used as a benchmark for evaluating a fine-grained code diversification technique.

\noindent
\label{exp:single-round-rand}
{\bf Methodology for single-round randomization experiments}.
In our experiments, we measure the occurrences of gadgets from the Turing-complete gadget set under fine-grained code randomization schemes. To enforce the code randomization schemes, we used four relatively new code randomization tools: Zipr~\cite{hawkins2017zipr} (instruction-level randomization), SR~\cite{conti2016selfrando} (function-level randomization), CCR~\cite{koo2018compiler} (block-level randomization), and MCR~\cite{homescu2013profile} (function + register-level randomization), because of their reliability. Table~\ref{tab:summary-rand-tools} in Appendix shows the key differences between these schemes. We compile and build a coarse- and a fine-grained version of each application or dynamic library for each run using each of the four randomization tools, i.e., each run has a different randomized code. We use LLVM Clang version 3.9.0, version 3.8.0 and GCC version 5.4.0 as the compilers for CCR, MCR and SR, respectively. We run, load or rewrite each application or library 100 times to reduce the impact of variability on the number of gadgets in each run or load.

We use ropper~\cite{ropper}, an offline gadget finder tool, under coarse-grained ASLR. Under fine-grained ASLR, we write a tool to recreate the JIT-ROP~\cite{snow2013just} exploitation process, including code page discovery and gadget mining. %We use capstone \cite{capstone-disas} for disassembling code pages. 
Our tool can search for gadgets of a specific type. We scan the opcodes of {\em ret} (0xC3) and {\em ret xxx} (0xC2) and perform a narrow-scoped backward disassembly from those locations to collect ROP gadgets. Similarly, we scan the opcodes of {\em int 0x80} (0xCD 0x80), {\em syscall} (0x0F 0x05), {\em sysenter} (0x0F 0x34) and {\em call gs:[10]} (0x65 xFF 0x15 0x10 0x00 0x00 0x00) for system gadgets. \color{black}We consider the gadgets only from the legitimate instructions, not from instructions within overlapping instruction bytes.\color{black} %We consider gadgets of length up to five (5) instructions, similar to the default setting in other ROP gadget finding tools (e.g., Ropper and ROPGadget).

\noindent
\label{exp:rerand-interval}
{\bf Methodology for re-randomization experiments}. For code re-randomization schemes, we attempted to use six re-randomization tools. However, some of the tools are unavailable and some have runtime and compile-time issues\footnote{Remix~\cite{chen2016remix} \& CodeArmor~\cite{chen2017codearmor} are not available. TASR~\cite{bigelow2015timely} is not accessible for policy issues. Runtime ASLR~\cite{lu2016make} \& Stabilizer~\cite{curtsinger2013stabilizer} have run \& compile time issues, respectively.}; 
in the end, we were able to obtain only Shuffler~\cite{williams2016shuffler}. To evaluate the impact of re-randomization, we take 100 consecutive address space snapshots from an application/library re-randomized by Shuffler~\cite{williams2016shuffler}. Then, we manually analyze the address space snapshots. 

The choice of re-randomization intervals is important for a re-randomization scheme. An effective re-randomization interval should hinder attackers' capabilities while ensuring performance guarantees. Our measurement methodology determines the upper bound (see definition~\ref{upperbound}) of effective re-randomization intervals by considering the fastest speed of gadget convergence, i.e., the minimum time for convergence. To measure the time of gadget convergence, we run the recursive code harvest process for an application and record the times it takes to converge to different gadget sets such as Turing-complete, priority, MOV TC, and payload gadget sets. We record the number of leaked gadget types that the code harvest process covered so far, while recording the convergence time. The code harvest terminates upon gadget convergence. \color{black}We record multiple convergence times by starting the code harvesting process from multiple pointer locations to capture the variability. To select multiple starting pointers, we choose a random code pointer from each code page of an application. Choosing a single random code pointer from each code page allows us to identify all instructions and pointers on that code page.\color{black}

%CCSR2020 The choice of re-randomization intervals is important for a re-randomization scheme. An effective re-randomization interval should hinder attackers' capabilities while ensuring performance guarantees. Our measurement methodology determines the upper bound (see definition \ref{upperbound}) of effective re-randomization intervals by considering the fastest speed of gadget convergence, i.e., the minimum time for convergence. To measure the time of gadget convergence, we run the dynamic code harvest process for an application. We record the time at the start of the code page harvest process, after each code page harvest, and the completion of the code page harvest process. We also determine the number of gadget types that the code harvest process covered so far while recording the time. The number of gadget types indicates convergence status. The code harvest process terminates upon gadget convergence. At this point, we can measure the time for the convergence from the start and end times. To measure the fastest speed of convergence, we run the dynamic code harvest process starting from a random code pointer from each code page of an application and record the minimum time required for the gadget convergence.

%For Shuffler~\cite{williams2016shuffler}, we do not need to prepare special binaries. 

%-----------------------------------------------------------------------------------
\subsection{\textbf{Methodology for System Access}} \label{exp-system-access}
%CCS20%\subsubsection{\textbf{Measurement Methodology for System Access}} \label{exp-system-access}
%\noindent
%{\bf Measurement Methodology for System Access}. \label{exp-system-access}
%This methodology is for evaluating our RQ \#2, and \#3 (see the evaluations in \S\ref{eval-RQ2} and \ref{eval-RQ3}, respectively). 
We measure the difficulty of accessing privileged operations through the availability of system gadgets and vulnerable library pointers in a stack, heap or data segment. For system gadgets, we compare the number of system gadgets under the coarse- and fine-grained randomization and compute the reduction in the gadget quantity.
%\noindent{\em System gadget measurement}: In addition to a system gadget (e.g., syscall), an adversary needs a set of pop gadgets (e.g., \verb1pop rax; ret;1) to setup the environment for a system call. In this part, we examine the impact of coarse- and fine-grained code randomization on the availability of these system call and related gadgets.
%
%
%In the presence of coarse-grain code randomization, many adversaries \cite{gil2018hole} leak known library pointers (e.g., libc function pointer) to derandomize the library. 
%
For the measurement of vulnerable pointers in a stack/heap/data-segment, we examine the overall risk associated with a stack/heap/data-segment by identifying the number of unique {\em libc} pointers in that stack/heap/data-segment. For the evaluation purpose, we do not exploit vulnerabilities to leak {\em libc} pointers from the stack/heap/data-segment. Rather, we assume that we know the address mapping of {\em libc} and can find the {\em libc} pointers through a linear scanning of the stack/heap/data-segment. We discuss the existence of {\em libc} pointers in popular applications in Section~\ref{evaluation-libc-pointers}.
%\noindent{\em Measurement of vulnerable pointers in stack/heap/data segment}: In the presence of coarse-grain code randomization, many adversaries \cite{gil2018hole} leak known library pointers (e.g., libc function pointer) to derandomize the library. For this measurement, we examine the overall risk associated with stack/heap/data-segment by identifying the number of unique libc pointers in stack/heap/data-segment. For evaluation purpose, we do not try to exploit vulnerabilities to leak libc pointers from stack/heap/data-segment, rather we assume that we know the address mapping of the libc library and can find the libc pointers through a linear scanning of stack/heap/data-segment. We discuss the existence of libc pointers in popular applications in \S\ref{eval-RQ5}.

%----------------------------------------------------------------------------------
\subsection{\textbf{Methodology for Payload Generation}} \label{exp-payload-gen}
%\noindent
%{\bf Measurement Methodology for Payload Generation}. \label{exp-payload-gen}
%CCS20% We evaluate our RQ \#3 (in Section~\ref{evaluation-payload-generation}) using this methodology.
%We focus on measuring the impact of fine-grained randomization on forming gadget chains. In particular, we look for the set of gadgets required for spawning a shell in both coarse- and fine-grained versions of an application/library. The required gadgets are Load Register (LR), Store Memory (SM), Move Register (MV) or XOR, Arithmetic (AM) or INC-DEC, and system (SYS) gadgets. LR loads a value from the top of a stack to register, SM performs write-what-where operation, and SYS gadget performs system calls. MV, XOR, AM, and INC-DEC gadgets are necessary for additions/subtractions in order to set up necessary register values for system calls. We measure the existence of these gadgets in both coarse- and fine-grained versions of an application/library to assess the impact of fine-grained randomization on forming gadget chains.
%
%
We focus on measuring the quality of individual gadgets to approximate the quality of a gadget chain. The quality of a set of gadgets for generating payloads is essential, as attackers need to use gadgets to set up and prepare register states.  To measure the quality of individual gadgets, we perform a register corruption analysis for each gadget, which is briefly described next.  The detail description of our register corruption analysis is in Appendix~\ref{appendix:register-corruption-analysis}.

Typically, a gadget contains one core instruction that serves the purpose of that gadget. For example, an MR gadget may contain {\em mov eax, edx} as the core instruction and some additional instructions before/after the core instruction. We measure the register corruption rate by analyzing how the core instruction of a gadget can get modified by those additional instructions. \color{black}In the case of multiple core instructions of a gadget type, we consider the core instruction that is closest to the ret instruction\color{black}. A core instruction may be modified by {\em i)} the instruction(s) before the core instruction, {\em ii)} the instruction(s) after the core instruction, and {\em iii)} both the instruction(s) before/after the core instruction. For each gadget, we consider these three scenarios and determine whether the gadget is corrupted or not. Next, we discuss the code randomization and re-randomization tools briefly in the following paragraphs.

{\it Shuffler}~\cite{williams2016shuffler} runs itself alongside the user space program that it aims to protect. It has a separate asynchronous thread that continuously permutes all the functions to make any memory leaks unusable as fast as possible.

{\it Zipr}~\cite{hawkins2017zipr} reorders the location of each instruction in an executable or library (an example in Figure~\ref{fig:ilr-protected-program} in the Appendix). Zipr works directly on binaries or libraries with no compiler supports. Zipr~\cite{hawkins2017zipr} is based on the Intermediate Representation Database (IRDB) code. Zipr shuffles code during the rewriting process, which is called block-level instruction layout randomization.

%{\em \textbf{Instruction Location Randomization (ILR)}}~\cite{hiser2012ilr} reorders the location of each instruction in an executable or library (Figure \ref{fig:ilr-protected-program}). ILR works directly on binaries or libraries with no compiler supports. Since ILR's implementation is no longer maintained, we use the Zipr~\cite{hawkins2017zipr} static rewriter. Zipr~\cite{hawkins2017zipr} is based on the Intermediate Representation (IR) Database (IRDB) code that was used for ILR. Zipr inherently shuffles code during the rewriting process, which is called block-level instruction layout randomization (BILR).

%CCSR2020 {\it Selfrando} (SR)~\cite{conti2016selfrando} applies code diversification at the load time. This tool collects \textit{Translation and Protection (TRaP)} information, a minimal set of metadata for function boundaries during the linking phase. This tool also inserts a dynamic library called {\em libselfrando}. At the load time, this library takes control of the execution, reorders the position of each function in an executable utilizing the TRaP information, and relinquishes the control to the original entry point of the executable. SR is compiler-agnostic and can use either GCC or Clang as its compilation engine.

{\it Selfrando} (SR)~\cite{conti2016selfrando} is compiler-agnostic and applies code diversification at the load time using function boundary-metadata called \textit{Translation and Protection (TRaP)} and inserting a dynamic library called {\em libselfrando}. At the load time, {\em libselfrando} takes control of the execution, reorders the position of each function in an executable utilizing the TRaP information, and relinquishes the control to the original entry point of the executable. %SR is compiler-agnostic and can use either GCC or Clang as its compilation engine.

{\it Multicompiler} (MCR)~\cite{homescu2013profile, crane2015readactor} applies the code diversification at the link time. This tool randomizes functions, machine registers, stack-layout, global symbols, VTable, PLT entries, and contents of the data section. The tool also supports insertion of NOP, global padding, and padding between stack frames. We choose the function and machine register level randomization for our evaluation. MCR uses the clang-3.8 LLVM compiler as its compilation engine.

{\it Compiler-Assisted Code Randomization} (CCR)~\cite{koo2018compiler} applies the code diversification at the installation time, i.e., rewrites an executable binary after reordering the functions and basic blocks of the executable. This tool collects metadata for code layout, block boundaries (i.e., the basic block, functional block, and object block boundaries), fixup, and jump table of an executable during compilation and linking phases. A Python script rewrites the executable binary utilizing the collected metadata.
%The tool embeds this metadata into the executable by adding a new section called {\em .rand}. A Python script then rewrites the executable binary by reordering the positions of basic and functional blocks. 
In our experiments, CCR uses the clang-3.9 LLVM compiler as its compilation engine.

%\noindent
\textit{Availability and robustness of fine-grained ASLR tools.}
We found that the majority of code diversification implementations, including ASR~\cite{giuffrida2012enhanced}, ASLP~\cite{kil2006address}, Remix~\cite{chen2016remix}, and STIR~\cite{wartell2012binary}, are not publicly available.
Some available tools (e.g., MCR~\cite{homescu2013profile, crane2015readactor}, CCR~\cite{koo2018compiler} and SR~\cite{conti2016selfrando}) operate on the source code level that requires recompilation. 
%They require the recompilation of source code including dynamic libraries. 
We experienced multiple linking issues while using CCR and SR to compile Glibc code. The tool authors confirmed the limitations (discussed in Section~\ref{glibc-diversification}). ORP~\cite{pappas2012smashing} was the randomization tool used in Snow {\it et al.}'s JIT-ROP demonstration~\cite{snow2013just}. It operates on Windows binaries, incompatible with our setup.

\section{Evaluation Results and Insights}\label{exp_eval}
%-------------------------------------------

%{\bf From Fabian, "To me, the most important takeaway is that all pointer leaks from an application’s code segment are equally viable. That statement is really not highlighted well enough with supporting data, and should be. It has implications for a number of defenses, but none of that is really discussed. Again, how the measurements impact our understanding of the space is not well presented."}

\noindent
\label{exp-setup}{\em \textbf{Experimental setup}.} 
We implemented a JIT-ROP native code module. All experiments are performed on a Linux machine with Ubuntu 16.04 LTS 64-bit operating system. We write Python and bash scripts for automating our analysis and measurement process. 
%The scripts have around 3,500 lines of code including around 2,000 lines of Python code on top of GDB-Python-Utils \cite{gdb-python-utilz}. We use the Python regular expression library \textit{re} for finding semantically different gadgets, e.g., the Turing-complete gadgets and attack-specific gadgets. To overcome the issues involved with JIT-ROP \cite{snow2013just} for searching gadgets on the fly at runtime, we run/load each application/library and attach the application/library to GDB. The Python scripts are also loadable in GDB. This setup allows us to avoid a process' memory mapping related complexities. 
\color{black}
We will publish our JIT-ROP native implementation, analysis tool and data.
\color{black}

%We perform our experiments on the latest and stable versions of \verb1bzip21, \verb1cherokee1, \verb1hiawatha1, \verb1httpd1, \verb1lighttpd1, \verb1mupdf1, \verb1nginx1, \verb1openssl1, \verb1proftpd1, \verb1sqlite1, \verb1openssh1, \verb1thttpd1, \verb1tor1, and \verb1pdf1. We also perform our experiments on dynamic libraries. Dynamic libraries include \verb1libcrypto1, \verb1libgmp1, \verb1libhogweed1, \verb1libxcb1, \verb1libpcre1, \verb1libgcrypt1, \verb1libgnutls1, \verb1libgpg-error1, \verb|libtasn1|, \verb1libz1, \verb1libnettle1, \verb1libopenjp21, \verb1libopenlibm1,  \verb|libpng16|,  \verb1libtomcrypt1, \verb1libunistring1, and \verb1libxml21.

We perform our experiments on the latest and stable versions of applications including {\em bzip2}, {\em cherokee}, {\em hiawatha}, {\em httpd}, {\em lighttpd}, {\em mupdf}, {\em nginx}, {\em openssl}, {\em proftpd}, {\em sqlite}, {\em openssh}, {\em thttpd}, {\em xpdf}, and {\em mupdf}, browsers including {\em firefox}, {\em chromium}\footnote{Due to the incompatibility of the LLVM compiler version and the use of custom linkers with custom linking flags, we are unable to randomize the Chromium browser using SR, CCR, and MCR. Zipr also fails to randomize chromium possibly due to the large size of the executable ($\sim$944MB). However, we include a non-randomized version of the chromium browser in our re-randomization experiments.}, {\em tor}, {\em midori}, {\em netsurf}, and {\em rekonq} and browser engines such as {\em webkit}. We also perform our experiments on dynamic libraries. Dynamic libraries include {\em libcrypto}, {\em libgmp}, {\em libhogweed}, {\em libxcb}, {\em libpcre}, {\em libgcrypt}, {\em libgnutls}, {\em libgpg-error}, {\em libtasn1}, {\em libz}, {\em libnettle}, {\em libopenjp2}, {\em libopenlibm},  {\em libpng16}, {\em libtomcrypt}, {\em libunistring}, {\em libxml2}, {\em libmozgtk}, {\em libmozsandbox}, {\em libxul},  {\em libmozsqlite3}, {\em liblgpllibs}, {\em libwebkit2gtk-3.0}, and {\em musl}. We select these applications or dynamic libraries by considering the fact that many attackers demonstrate their attacks on most of these applications or libraries. Besides, these applications/libraries include a diverse set of areas such as the web server, browser, PDF reader, networking, database, and libraries in cryptography, math, image, and system.

\begin{table}[ht!]
%\footnotesize
\scriptsize
\caption{Numbers of the applications and dynamic libraries for experiments.}
\label{tab:app-lib-count}
\begin{tabular}{lcc}
\hline
Experiment & \begin{tabular}[c]{@{}c@{}}Applications (20 Total)\end{tabular} & \begin{tabular}[c]{@{}c@{}}Libraries (25 Total)\end{tabular} \\ \hline
Re-randomization interval & 17 & 15 \\
Instruction-level rand. & 15 & 14 \\
Function-level rand. & 17 & 21 \\
Function + register-level rand. & 12 & 13 \\
Basic block-level rand. & 15 & 15 \\ \hline
\end{tabular}
\end{table}

Table~\ref{tab:app-lib-count} shows the numbers of applications/libraries used for measuring the upper bound for re-randomization intervals and evaluating instruction-level~\cite{hawkins2017zipr}, functional-level~\cite{conti2016selfrando}, function+register-level~\cite{homescu2013profile, crane2015readactor}, and basic block-level~\cite{koo2018compiler} randomizations. Each experiment evaluates a different set of applications and libraries because no (re-)randomization tool is capable of (re-)randomizing all of our selected applications (20 in total) and libraries (25 in total). However, we also conduct our experiments and report results using the common set of applications and libraries.

%Table \ref{tab:app-lib-count} shows the numbers of applications/libraries used for evaluating Shuffler~\cite{williams2016shuffler}, Zipr~\cite{hawkins2017zipr}, SR~\cite{conti2016selfrando}, CCR~\cite{koo2018compiler}, and MCR~\cite{homescu2013profile}. Each tool evaluates a different set of applications and libraries because no tool is capable of (re-)randomizing all of our selected applications (13 in total) and libraries (19 in total). However, we also evaluate these tools using the common set of applications and libraries that these tools can randomize. 

%To evaluate the impact of re-randomization, we take 100 consecutive address space snapshots from an application/library re-randomized by Shuffler~\cite{williams2016shuffler}. Then, we manually analyze the address space snapshots. 
%
%We manually compile (or rewrite the executables of) the programs above to enforce fine-grained code randomization up to function level using SR~\cite{conti2016selfrando}, basic block level using CCR~\cite{koo2018compiler}, both functional and register levels using MCR~\cite{homescu2013profile}, and instruction level using Zipr~\cite{hawkins2017zipr}. We use LLVM Clang version 3.9.0, version 3.8.0 and GCC version 5.4.0 as the compilers for CCR, MCR and SR, respectively. We run, load or rewrite each application or dynamic library 100 times to reduce the impact of variability on the number of gadgets in each run or load.

We measure a total of 11 types of gadgets for the Turing-complete set, 10 types for the priority set, and 7 types for the MOV TC set. Different payloads have different types and numbers of gadgets.
%CCSR202 We measure a total of 15 types of gadgets, including 11 gadgets for Turing-complete operations and 4 attack specific gadgets for assessing the impact of randomization schemes. A set of gadget is {\em Turing-incomplete} if the set of gadgets does not include all 11 gadget types required for all Turing-complete operations.

%------------------------------------------------
%\subsection{Experimental Setup} \label{exp-setup}
%------------------------------------------------

%Our main goal is to quantitatively measure how and to what extent a fine-grained code randomization technique impacts on the availability of gadgets, specially the Turing-complete gadgets and some attack-specific gadgets. 

%The main goal of this study to understand to what extent code randomization and re-randomization solutions harden JIT-ROP attacks. The understanding of the capabilities and limitations of current (re-)randomization solutions will help (re-)randomization designers to design even better techniques.

%\color{blue}
\subsection{Re-randomization Upper Bound}
\color{black}
We determine the upper bound of re-randomization intervals by measuring the fastest speed of gadget convergence across the Turing-complete, priority, MOV TC, and payload gadget sets, i.e., measuring the minimum time that an attacker needs to collect all gadget types from any of the four gadget sets. Table~\ref{tab:summary-rerand-timing} shows the minimum (i.e., fastest speed) and the average time to leak all gadget types in a set. The minimum and the average time is calculated over 17 applications/browsers. From the table, we notice that the re-randomization upper bounds, i.e., the minimum time, range from 1.5 to 3.5 seconds. We observe some variability ($\sigma=0.8$) in the minimum time with the priority and MOV TC gadget sets having the lowest (1.5s) and highest (3.5s) time, respectively. Intuitively, the reason for this variability could be related to the number of gadget types necessary for each gadget set. However, we observe that the minimum time for the MOV TC gadget set is larger than the TC or priority gadget set even though the MOV TC has fewer gadget types. To understand more about this variability, we analyze how gadget types are leaked over time for individual applications/browsers across the four types of gadget sets.

\begin{table}[!ht]
\scriptsize
\caption{\textcolor{black}{Minimum and average time to leak all gadget types from TC, priority, MOV TC, and payload gadget sets. The percentage (\%) of time is spent for leaking gadgets versus analyzing gadgets. The minimum, average, and percentages for each set are calculated using 17 applications/browsers. Payload* $\longrightarrow$ average of three payload sets.}}
\label{tab:summary-rerand-timing}
\begin{tabular}{cggcc}
%\hline
 & \multicolumn{2}{g}{\begin{tabular}[c]{@{}c@{}}Time to leak all gadget types\end{tabular}} & \multicolumn{2}{c}{Gadget analysis} \\ \hline
\textbf{Gadget set} & \textbf{Minimum (s)} & \textbf{Average (s) } & \textbf{Leak (\%)} & \textbf{Analysis (\%)} \\ \hline
TC & 2.2 & 4.3 & 17 & 83 \\ %\hline
Priority & 1.5 & 3.5 & 13 & 87 \\ %\hline
MOV TC & 3.5 & 5.3 & 16 & 84 \\ %\hline
Payload* & 2.1 & 4.8 & 12 & 88 \\ \hline
Average & \textbf{2.3s} & \textbf{4.5s} & \textbf{14.5\%} & \textbf{85.5\%} \\ %\hline
\end{tabular}
\end{table}

\begin{figure}[!bth]
  \centering
  \includegraphics[width=1.0\linewidth]{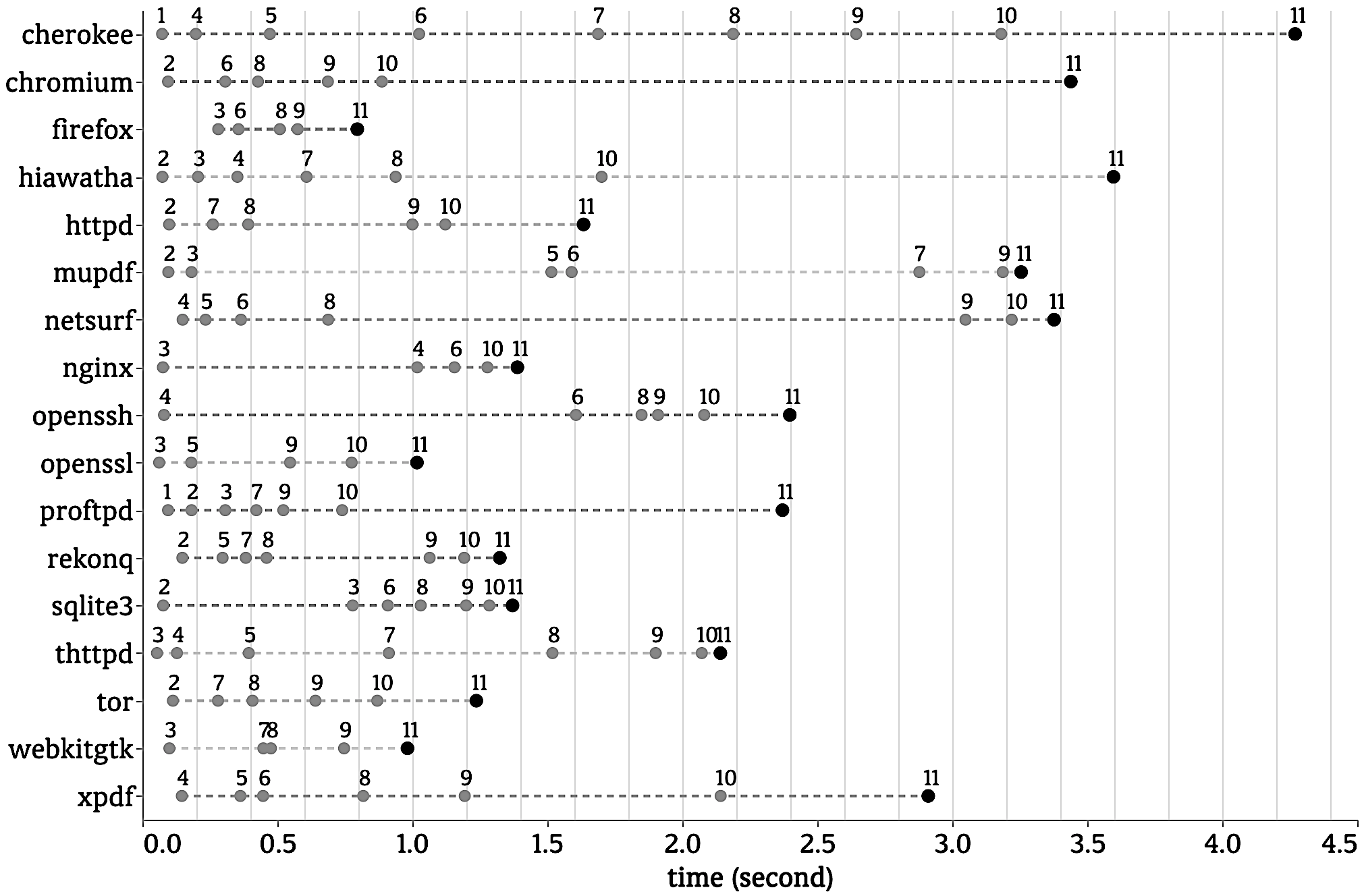}
  \caption{\textcolor{black}{Minimum time to obtain the Turing-complete gadget set with a timeline for new gadget type leaks. Each gray filled circle (\tikzcircle{2pt}) with a number $n$ on top of it represents the time to leak $n$ gadget types. The bold filled circle (\tikzcircle[black, fill=black]{2pt}) indicates the time to leak all gadget types. Applications and browsers are randomized with a function-level scheme~\cite{conti2016selfrando}.}}
  %\caption{\textcolor{blue}{Gadget convergence with the number of gadget types leaked over time. The \tikzcircle{2pt} mark indicates a certain time in a trajectory line when new gadget types are discovered. The number on top of the \tikzcircle{2pt} mark represents the number of leaked gadget types at a certain time. The time indicated by the bold \tikzcircle[black, fill=black]{2pt} mark is the re-randomization upper bound for an application. Applications and browsers are randomized with a function-level scheme~\cite{conti2016selfrando}.}}
  \label{fig:rerand-plot}
\end{figure}

Figure~\ref{fig:rerand-plot} shows the minimum time to obtain the Turing-complete gadget set from individual application/browser along with a timeline for new gadget type leaks. Each gray \tikzcircle{2pt} mark with a number $n$ on top of it represents the time to leak $n$ gadget types.  The bold \tikzcircle[black, fill=black]{2pt} mark represents the time to leak 11 gadget types from the Turing-complete gadget set. For example, it takes roughly 1 and 4.3 seconds to leak 6 and 11 gadget types, respectively from {\em cherokee}.

The number of leaks increases as time increases. However, the effect of the increase may not be immediate. For example, in Figure~\ref{fig:rerand-plot}, the code harvest process takes roughly 0.7 seconds to leak 8 distinct gadget types from {\em netsurf}. If the time increases to 1 or 2 seconds, the number of leaked gadgets is still the same, i.e., 8 distinct gadget types. However, if the time is more than 3 seconds, the number of leaked gadgets starts to increase. We call the time between 0.7 to 3 seconds as non-reactive.

\begin{table*}[!ht]
\centering
%\footnotesize
\scriptsize
\caption{\textcolor{black}{Impact of locations of pointer leaks on gadget availability.} The same application has different numbers of leaked addresses for different tools because each tool uses a different backend (i.e., compiler). Different backends produce different sized executables of the same program. Size of an executable is proportional to the number of code pages. Also, the numbers of gadgets from the \textcolor{black}{Function-level scheme~\cite{conti2016selfrando} and Function + register-level scheme~\cite{homescu2013profile, crane2015readactor} are not comparable due to their different backends.}}
\label{tab:location-pointer-leak}

\renewcommand{\arraystretch}{1.1}
\setlength\tabcolsep{5pt} % default value: 6pt

\begin{tabular}{lggglllggglll}
 & \multicolumn{3}{c}{Instruction-level scheme~\cite{hawkins2017zipr}} & \multicolumn{3}{c}{Function-level scheme~\cite{conti2016selfrando}} & \multicolumn{3}{c}{Function + register-level scheme~\cite{homescu2013profile, crane2015readactor}} & \multicolumn{3}{c}{Block-level scheme~\cite{koo2018compiler}} \\ 
 \hline
%\rowcolor{backgroundColour}
Program & \begin{tabular}[c]{@{}l@{}}\# of leaked \\ addresses\end{tabular} & \begin{tabular}[c]{@{}l@{}}\# of \\ MIN-FP\end{tabular} & \begin{tabular}[c]{@{}l@{}}\# of \\ EX-FP\end{tabular} & \begin{tabular}[c]{@{}l@{}}\# of leaked\\ addresses\end{tabular} & \begin{tabular}[c]{@{}l@{}}\# of\\ MIN-FP\end{tabular} & \begin{tabular}[c]{@{}l@{}}\# of\\ EX-FP\end{tabular} & \begin{tabular}[c]{@{}l@{}}\# of leaked\\ addresses\end{tabular} & \begin{tabular}[c]{@{}l@{}}\# of\\ MIN-FP\end{tabular} & \begin{tabular}[c]{@{}l@{}}\# of\\ EX-FP\end{tabular} & \begin{tabular}[c]{@{}l@{}}\# of leaked\\ addresses\end{tabular} & \begin{tabular}[c]{@{}l@{}}\# of\\ MIN-FP\end{tabular} & \begin{tabular}[c]{@{}l@{}}\# of\\ EX-FP\end{tabular} \\ 
\hline
hiawatha & 41 & 9 & 223 & 42 & 41 & 1259 & 47 & 44 & 1042 & 39 & 31 & 793 \\
httpd & 91 & 16 & 634 & 91 & 141 & 4453 & \multicolumn{3}{g}{MCR produces linking error for httpd} & 86 & 176 & 4764 \\
lighttpd & 53 & 8 & 235 & 53 & 103 & 2512 & 68 & 118 & 2544 & 45 & 74 & 1783 \\
nginx & 114 & 26 & 788 & 121 & 222 & 5277 & 49 & 111 & 1731 & 114 & 204 & 4822 \\
proftpd & 131 & 17 & 523 & 187 & 96 & 7395 & 131 & 115 & 4466 & 131 & 125 & 3986 \\
thttpd & 10 & 8 & 172 & 17 & 22 & 583 & 16 & 31 & 535 & 15 & 24 & 428
\end{tabular}
\end{table*}

We observe a number of long non-reactive times for some other applications such as chromium (0.89--3.44s), hiawatha (1.7--3.6s), mupdf (0.18--1.52s and 1.6--2.88s), openssh (0.08--1.61s), proftpd (0.74--2.37s), and xpdf (1.19--2.14s). Most of these non-reactive times are towards the end of their timelines. These non-reactive times indicate that a few missing gadget types prevent the discovered set from being Turing-complete quickly. That is, a few types of gadgets are very scarce. The scarcest gadgets are Load-Memory (LM), Arithmetic-Load (AM-LD), and System Call (SYS) gadgets. 
%For example, the gadget discovery process searches for the Load-Memory (LR) gadgets for around 7 seconds for proftpd. 
The fundamental reason for the scarcity is that some applications (including libraries) have a few register-based memory accesses. %The memory accesses are made by adding some offsets to the registers. 
Besides, the main executable of an application does not have SYS gadgets in most cases.    
%\color{black}

We also observe similar non-reactive times for obtaining the priority and MOV TC gadget sets. The variability in the minimum time of the four gadget sets is due to the Arithmetic-Load (AM-LD) gadget type. Since the priority gadget set does not include \texttt{AM-LD}, its code page harvest process is the fastest. The time for the MOV TC gadget set is relatively longer than the TC and priority gadget set, even though MOV TC does not include \texttt{AM-LD}. The reason for this long time is that the MOV TC set includes several specialized Load-Memory (LM) and Store-Memory (ST) gadget types.

The MOV TC gadget set is powerful since it takes only a few {\em mov} instructions with four register pairs to perform the Turing-complete operations. To observe to what extent MOV TC gadgets are prevalent in applications, we count the numbers of six MOV gadgets (\texttt{MR}, \texttt{ST}, \texttt{STCONSTEX}, \texttt{STCONST}, \texttt{LM}, and \texttt{LMEX} described Table~\ref{tab:gadgets-priority-movtc-sets} in the Appendix) and the System Call (\texttt{SYS}) gadget while measuring the minimum time to find these gadgets. \texttt{STCONSTEX}, \texttt{STCONST}, and \texttt{LMEX} gadgets are variants of \texttt{ST} and \texttt{LM} gadgets. The average number of gadgets for \texttt{MR} is 51, \texttt{ST} is 14, \texttt{STCONSTEX} is 35, \texttt{STCONST} is 2, \texttt{LM} is 3, and \texttt{LMEX} is 15. As expected, the number of Load-Memory (\texttt{LM}) gadgets is low, which indicates the scarcity of this gadgets. Besides, we observe the number of Store-Constant (\texttt{STCONST}) is also low, which is necessary for performing comparison and conditional operations. The average number of (\texttt{SYS}) gadgets is adequate (23).

Our re-randomization upper bound calculation includes the overhead of analyzing different gadget types. Thus, we perform additional analyses to investigate how fast the address space is leaked and how much time is spent on gadget analysis. According to our findings (Table~\ref{tab:summary-rerand-timing}), we find that on average around {\bf 15\%} of the time is spent on leaking address space, while the rest for gadget searching. This result indicates that a JIT-ROP attacker spends a significant amount of time searching for gadget types. Thus, the upper bound of re-randomization intervals is subject to change based upon an optimized gadget search strategy. %Additionally, the amount of code discovered correlate with the number of different gadget types found?

\color{black}

Clearly, the value of the upper bound for the re-randomization intervals also depends on the machine (e.g., CPUs, cache size, memory, etc.) where the measurement is conducted. %The intervals for obtaining the Turing-complete gadget set are also different for different applications.
%Figure \ref{rerand-plot} shows that the lowest re-randomization interval to achieve 100\% of Turing-complete gadgets is around 4 seconds in our experiments. The number is computed by taking the lowest value that is observed across the 11 applications (with the associated dynamic libraries) in Figure \ref{rerand-plot} to achieve the Turing-complete gadget set. We choose the lowest value to ensure security guarantees. 
Using our methodology, defenders can perform the measurement on their machines to determine what intervals are appropriate for their applications, while satisfying overhead constraints. In Section~\ref{sec:discussion}, we discuss the implications of re-randomization intervals in real-world operations.%, e.g., how to choose re-randomization intervals for performance- and security-critical applications. 

%\color{blue}
We call the upper bound of re-randomization intervals as the ``best-case'' re-randomization interval from a defender's perspective because the defender has to rerandomize by the time of the interval, if not sooner. This raises the question regarding the effectiveness of ``best-case'' intervals over ``worst-case'' intervals. The ``worst-case'' interval indicates the time required to build a useful gadget chain using a minimal set of gadgets. In reality, attackers' goals vary. It is difficult to determine a minimum set of gadgets common and necessary across all attack chains. %Besides, a minimal set could prevent a set of gadgets from obtaining the Turing-complete expressive power of ROP.
Our ``best-case'' interval includes the time for discovering SYS gadgets that are scarce. Some attack scenarios may not require the SYS gadgets, but the necessity of SYS gadgets or system APIs in attack chains have been shown by previous work~\cite{bletsch2011jump, bittau2014hacking, snow2013just, carlini2014rop, davi2015isomeron}.
%\color{black}

%--------------------------------------------------------------------
\subsection{Impact of the Location of Pointer Leakage}\label{eval-impact-location-pointer}
%--------------------------------------------------------------------
%CCS20% \noindent \textbf{\textit{RQ \#2}}: How does the location of a code pointer leak impact the availability of gadgets in the presence of fine-grained code randomization?
We measure the impact of pointer locations on JIT-ROP attack capabilities, by comparing the number of gadgets harvested and the time of harvest under different {\em starting} pointer locations. We aim to find out whether or not the number of gadgets and the time depends on the location of a pointer leakage when a fine-grained randomization scheme is applied.

%CCS20% We measure the impact of pointer locations on JIT-ROP attack capabilities, by comparing the number of gadgets harvested under different {\em starting} pointer locations. We aim to find out whether or not the number of gadgets depends on the location of a pointer leakage when a fine-grained diversification is applied.
%
%
%We leak a random code pointer from each code page. The pointer is the starting location. Since we harvest code page by code page, a leak in any location of a code page allows us to read the whole code page (4096 bytes). We discover new code pages using the recursive code page harvest technique from the starting location. We disassemble code pages, identify total gadgets, and extract semantically different gadget types.
%
%In Figure \ref{location-impact-gadgets}, each data point represents a total number of gadgets dynamically harvested from code pages by starting from a leaked pointer. The X-axis represents different locations of the leaked pointer. The Y-axis is the total number of gadgets. We evaluated four programs and the Libc library.
%

\noindent
\textbf{Impact of pointer locations on gadget availability}. 
To measure the impact of pointer locations on gadget availability, we collect the number of minimum and extended footprint gadgets by leaking a random code pointer from {\bf each} code page of {\em hiawatha}, {\em httpd}, {\em lighttpd}, {\em nginx}, {\em proftpd}, and {\em thttpd} and starting the code harvesting process from that leaked code pointer. Then we calculate the average number of gadgets for each leaked pointer. We leak a single code pointer from a single code page randomly because choosing any single random code pointer from a code page allows us to identify all instructions and all code pointers on that code page. Table~\ref{tab:location-pointer-leak} shows the number of leak code pointers or addresses and the numbers of minimum and extended footprint gadgets. We restrict the code harvest process to harvest gadgets from the main executable of an application to find how well the code of that application is connected. \color{black}We exclude the dynamic libraries for this experiment because many applications use a common set of libraries and the gadgets from this common set of libraries (if not excluded) would dominate the total number of gadgets.
\color{black}

%To measure the impact of pointer locations on gadget availability, we collect the total numbers of minimum and extended footprint gadgets by leaking a random code pointer from {\bf each} code page of {\em hiawatha}, {\em httpd}, {\em lighttpd}, {\em nginx}, {\em proftpd}, and {\em thttpd}. Choosing a single random code pointer from each code page allows us to identify all instructions on that code page. Table \ref{tab:location-pointer-leak} shows the number of leak code pointers or addresses and the numbers of minimum and extended footprint gadgets that can be harvested by starting from the leaked pointers. We restrict the code harvest process to harvest gadgets from the main executable of an application to find how well the code of that application is connected. 

For all applications, we observe that the pointer's location does not have any impact on the total number of minimum and extended footprint gadgets. For example, regardless of the location of starting point in nginx, we observe 26 minimum and 788 extended gadgets when randomized by the instruction-level randomization scheme; 222 minimum and 5277 extended footprint gadgets when randomized by the function-level scheme; 111 minimum and 1731 extended footprint gadgets when randomized by function + register-level scheme; and 204 minimum and 4822 extended footprint gadgets when randomized by block-level scheme. {\bf These findings indicate that an application's code segment is very well-connected, making JIT-ROP attacks easier}.

%CCS20% We collect the total numbers of minimum and extended footprint gadgets by leaking a random code pointer from {\bf each} code page of hiawatha, httpd, lighttpd, nginx, proftpd, and thttpd. %Each application is randomized using ILR, SR, MCR, and CCR. 

%CCS20% Table \ref{tab:location-pointer-leak} shows the number of leak code pointers or addresses and the numbers of minimum and extended footprint gadgets that can be harvested by starting from the leaked pointers. We restrict the code harvesting process to harvest gadgets only within the text segment of an application to find how well the code of a program is connected. For all applications, we observe that \textbf{the pointer's location does not have any impact} on the total number of minimum and extended footprint gadgets. For example, regardless of the location of starting point in nginx, we observe 26 minimum and 788 extended gadgets when randomized by the instruction-level randomization scheme; 222 minimum and 5277 extended footprint gadgets when randomized by the function-level scheme; 111 minimum and 1731 extended footprint gadgets when randomized by function + register-level scheme; and 204 minimum and 4822 extended footprint gadgets when randomized by block-level scheme. These findings indicate that an application's code segment is very well-connected, making JIT-ROP attacks easier.

The numbers of leaked addresses in Table~\ref{tab:location-pointer-leak} are different for different randomization schemes because we use different tools to enforce these randomization schemes. Different tools use different backends and different backends optimize the same application differently. This increases/decreases the number of code pages. Since we leak a random address from each code page, the number of leaked addresses varies with tools.

\begin{figure}[!h]
    \centering
    \includegraphics[width=0.48\textwidth]{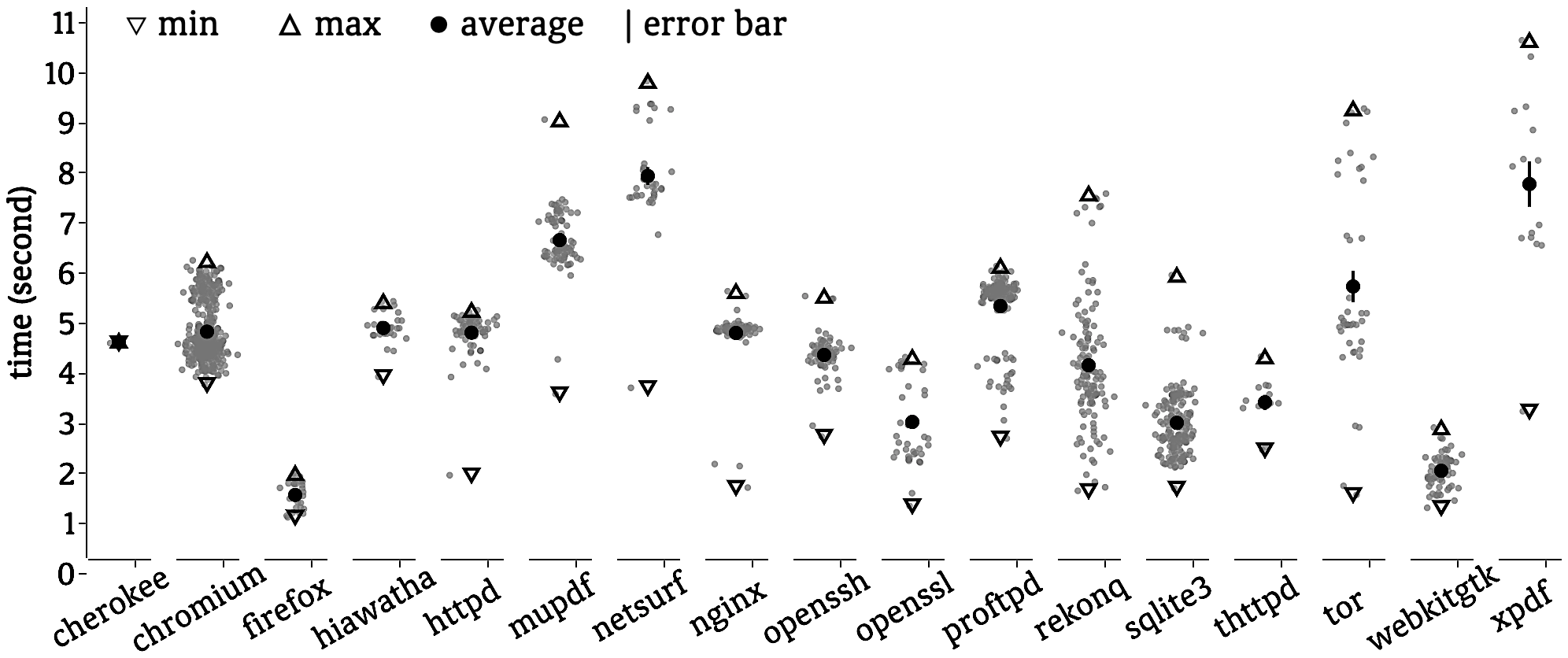}
    \caption{\textcolor{black}{Impact of starting pointer locations on gadget harvesting time. Each \tikzcircle{2pt} indicates the time for harvesting the Turing-complete gadget set. The minimum, maximum, and average time is calculated by running code harvest process from multiple starting code pointer locations. A small amount of jitter has been added to x-axis for each application/browser for better visibility of times along the y-axis.}}
    %\caption{Times needed to harvest the Turing-complete gadget set. The minimum, maximum, and average times are calculated from the times that the code harvest process takes by starting from a code pointer leaked from each code page.}
    \label{fig:time-for-gadget-convergence}
\end{figure}

%\color{blue}
\textbf{Impact of pointer locations on code harvest time}. To measure the impact of code pointer locations on the time, we measure the time required to leak all gadget types from the Turing-complete gadget set. We run the code harvest process starting from a random code pointer leaked from each code page of an application or browser and record the time to collect all gadget types. Figure~\ref{fig:time-for-gadget-convergence} shows the minimum, maximum, and average time to leak all gadgets for different applications and browsers. For a few code pointers from several applications/browsers (e.g., 3 out of 111 code pointers for {\em nginx} or 8 out of 40 code pointers for {\em openssl} or 2 out of 41 for {\em tor}), the code harvest process takes significantly shorter time than the average. We analyze the reason for this phenomenon.

We find that most applications/browsers have some code pages that contain a diverse set of gadgets. For example, {\em nginx} contains 9 code pages that have at least 5 distinct gadget types from the Turing-complete gadget set. Whenever the code harvest process accesses those code pages sooner, the discovered gadgets quickly converge to Turing-complete.

\noindent\textit{Future directions.} %CCS20% Our findings imply that \textbf{any} valid code pointer leak from an application's code segment is equally viable. Regardless of randomization, a pointer leakage in any location allows attackers to access a set of minimum and extended footprint gadgets. These observations reassert that disrupting the connectivity of the code segment is an effective defense strategy. Security tools including but not limited to Oxymoron \cite{backes2014oxymoron}, Readactor~\cite{crane2015readactor}, XnR~\cite{backes2014you}, NEAR~\cite{werner2016no}, Heisenbyte~\cite{tang2015heisenbyte}, and ASLR-Guard~\cite{lu2015aslr} specifically disrupt code connectivity and prevent pointer leakage. A large-scale quantitative assessment on the effectiveness of these security tools is necessary to find out the practicality and feasibility of these tools for deployment. Also, the design of risk heuristics-based pointer selection and prioritization for protecting pointers from leakage would be an interesting direction.
Our findings imply that any valid code pointer leak is equally viable with regards to the coverage of gadgets. This observation reasserts that disrupting the code connectivity is an effective defense strategy used in Oxymoron~\cite{backes2014oxymoron}, Readactor~\cite{crane2015readactor}, XnR~\cite{backes2014you}, NEAR~\cite{werner2016no}, Heisenbyte~\cite{tang2015heisenbyte}, and ASLR-Guard~\cite{lu2015aslr} tools. A large-scale quantitative assessment on the effectiveness of these security tools is necessary to find out the practicality and feasibility of these tools for deployment. \color{black}Also, the design of risk heuristics-based pointer selection and prioritization for protecting pointers from leakage would be an interesting direction. The idea is to prioritize code pointers based on the convergence time and data pointers based on their sensitivity (e.g., data pointers used in loops). %Then these prioritized pointers are protected from leakage.
\color{black}

%Also, the design of risk heuristics-based pointer selection and prioritization by considering the two metrics (i.e., gadget convergence and time for the convergence) for protecting pointers from leakage would be an interesting direction.

%Although this kind of disruption solutions (e.g., Oxymoron \cite{backes2014oxymoron}) exist, they increase the runtime overhead and cannot protect from the variants of JIT-ROP (e.g., Isomeron~\cite{davi2015isomeron}). Thus, a randomization time solution that disrupts the connectivity of code while keeping the execution order intact would be an interesting research direction.

%\noindent\textbf{Summary}: The location of the pointer leak does not have an impact on the gadget availability in JIT-ROP. Regardless of whether coarse- or fine-grained randomization is in place, a pointer leakage in any location allows attackers to access a set of minimum footprint and extended footprint gadgets.

%------------------------------------------------------------------------------------------------------
\subsection{Impact on the Availability of Gadgets}\label{eval-RQ1}
\textbf{\textit{Impact of Single-round Randomization Schemes}}.
Table~\ref{tab:summary-main} summarizes the impact of fine-grained code randomization schemes on the availability of gadgets in various applications (i.e., the main executables) and dynamic libraries. We measure the numbers of the various gadgets (as mentioned in Section~\ref{exp:single-round-rand}) for each application and library before and after randomizing with the four fine-grained randomization schemes. % (i.e., Zipr~\cite{hawkins2017zipr}, SR~\cite{conti2016selfrando}, MCR~\cite{homescu2013profile}, and CCR~\cite{koo2018compiler}).
Each application or library is run/loaded in memory for 100 times after randomizing 100 times when necessary\footnote{One compilation with 100 runs, 100 times randomization, 100 times compilation, and 100 times rewriting are required for SR, CCR, MCR, and Zipr, respectively.}. The numbers of gadgets are averaged over 100 runs/loads of an application or library. Then the numbers of gadgets are averaged over the number of applications and libraries for each randomization scheme. Table~\ref{tab:summary-main} shows the overall gadget reductions in application and library categories for each randomization scheme.%Zipr~\cite{hawkins2017zipr}, SR~\cite{conti2016selfrando}, MCR~\cite{homescu2013profile}, and CCR~\cite{koo2018compiler}. %Again, by reduction, we mean that these randomization tools thwart the current finding tools to obtain gadgets from an application's address space.

On average, the number of gadgets is reduced (by 18\%--28\% for minimum footprint and 37\%--45\% for extended footprint gadgets) when applications are randomized using function-, block-, and function+register-level schemes. % SR, CCR, and MCR. 
For dynamic libraries, the reductions range from around 21\%--47\% for minimum footprint gadgets and around 37\%--44\% for extended footprint gadgets. However, instruction-level randomization scheme %Zipr~\cite{hawkins2017zipr} 
reduces the overall gadget amount significantly by around 80\%--90\% for both minimum and extended footprint gadgets. Table~\ref{tab:summary-main} also shows the reduction of gadgets in seven Turing-complete (TC) operations and indicates whether the Turing-complete expressiveness is preserved after applying the code randomization. The numbers before and after a vertical bar (\textbar) indicate the reduction of minimum and extended footprint gadgets for a Turing-complete operation. Since the numbers of applications/libraries are different for randomization schemes, we also evaluate these schemes and validate the results (Figure~\ref{fig:reduction-same-set} in the Appendix) with a common set of applications and libraries. The results show a consistent reduction in all the schemes. %Figure \ref{fig:reduction-same-set} in the Appendix shows the validation results. According to the results, all the schemes exhibit a consistent reduction while we evaluate them using the common set of applications and libraries.

\begin{table*}[!ht]
\centering
\scriptsize
%\small
%\footnotesize
\caption{Impact of fine-grained single-round randomization on the availability of gadgets in various applications and dynamic libraries. Instruction-level randomization scheme~\cite{hawkins2017zipr} is applied on 15 applications and 14 dynamic libraries, function-level scheme~\cite{conti2016selfrando} on 17 applications and 21 dynamic libraries, function + register-level scheme~\cite{homescu2013profile, crane2015readactor} on 12 applications and 13 dynamic libraries, and basic block-level scheme~\cite{koo2018compiler} on 15 applications and 15 dynamic libraries. The data of each application or library is the average result of 100 runs/loads/rewrites. The standard deviations vary between 0.3$\sim$3.4 for minimum footprint and 5.04$\sim$22.85 for extended footprint gadgets. $\Downarrow$ indicates reduction.}
%\caption{Impact of fine-grained single-round randomization on the availability of gadgets in various applications and dynamic libraries. The data in each row of SR~\cite{conti2016selfrando} is generated by averaging the data of 11 applications and 15 dynamic libraries. The data in each row of Zipr~\cite{hawkins2017zipr} and CCR~\cite{koo2018compiler} is generated by averaging 9 applications and 9 libraries for each tool. 8 applications and 8 dynamic libraries are used for MCR~\cite{homescu2013profile}. The data of each application or library is the average result of 100 runs/loads/rewrites. The standard deviations vary between 0.3$\sim$3.4 for minimum footprint and 5.04$\sim$22.85 for extended footprint gadgets. $\Downarrow$ indicates reduction.}
\label{tab:summary-main}

\renewcommand{\arraystretch}{1.3}
\setlength\tabcolsep{5pt} % default value: 6pt

\begin{tabular}{lllllllllllc}
%\cline{6-12}
 &  &  & \multicolumn{1}{l}{} & \multicolumn{7}{c}{Reduction (\%) of Turing-complete (TC) gadgets in 7 TC operations (MIN-FP \textbar\ EX-FP)} & \\ \hline
 \rowcolor{backgroundColour}
\multicolumn{1}{l}{Randomization schemes} & \multicolumn{1}{l}{Granularity} & \multicolumn{1}{l}{\begin{tabular}[c]{@{}l@{}}$\Downarrow$ (\%)\\MIN-FP \end{tabular}} & \multicolumn{1}{l}{\begin{tabular}[c]{@{}l@{}}$\Downarrow$ (\%)\\EX-FP\end{tabular}} & \multicolumn{1}{|l}{Memory} & \multicolumn{1}{l}{Assignment} & \multicolumn{1}{l}{Arithmetic} & \multicolumn{1}{l}{Logical} & \multicolumn{1}{l}{\begin{tabular}[c]{@{}c@{}}Control\\ Flow\end{tabular}} & \multicolumn{1}{l}{\begin{tabular}[c]{@{}c@{}}Function\\ Call\end{tabular}} & \multicolumn{1}{l|}{\begin{tabular}[c]{@{}c@{}}System\\ Call\end{tabular}} & \multicolumn{1}{l}{\begin{tabular}[c]{@{}c@{}}TC\\ Preserved?\end{tabular}} \\ \hline
\multicolumn{12}{c}{\textbf{Applications}} \\ \hline
Inst. level rando.~\cite{hawkins2017zipr} & Inst. & \textbf{79.7} & \textbf{82.5} & \textbf{97.4} \textbar\ 82.7 & 58.8 \textbar\ 81.7 & \textbf{95.9} \textbar\ 64.9 & 85.8 \textbar\ 85.4 & 49.4 \textbar\ 80.1 & 67.4 \textbar\ 83.9 & 83.3 \textbar\ 0 & \xmark*  \\ 
Func. level rando.~\cite{conti2016selfrando}  & FB & 27.63 & 36.55 & 0.8 \textbar\ 29.2 & 10.6 \textbar\ 43.5 & 19.3 \textbar\ 15.1 & 35.1 \textbar\ 35.9 & 21.1 \textbar\ 29.1 & 18.2 \textbar\ 46.9 & 0 \textbar\ 0 & \cmark \\
Func.+Reg. level rando.~\cite{homescu2013profile} & FB \& Reg. & 17.62 & 42.37 & -8.3 \textbar\ 35.0 & -5.1 \textbar\ 35.2 & 26.1 \textbar\ 44.9 & 21.3 \textbar\ 38.1 & 34.0 \textbar\ 60.2 & 11.8 \textbar\ 64.9 & 80.0 \textbar\ 0 & \cmark  \\ 
Block level rand.~\cite{koo2018compiler} & BB & 19.58 & 44.64 & 5.5 \textbar\ 40.9 & 6.1 \textbar\ 47 & 26.1 \textbar\ 33.7 & 20.4 \textbar\ 37.4 & 41.2 \textbar\ 63.1 & 23.3 \textbar\ 56.3 & 0.0 \textbar\ 0 & \cmark  \\ \hline
\multicolumn{12}{c}{\textbf{Libraries}} \\ \hline
Inst. level rando.~\cite{hawkins2017zipr} & Inst. & \textbf{81.3} & \textbf{92.2} & \textbf{93.7} \textbar\ 96.1 & 60.7 \textbar\ 93 & \textbf{91.8} \textbar\ 84.9 & 84.5 \textbar\ 90.4 & 59.8 \textbar\ 93.5 & 51.8 \textbar\ 92.9 & 66.7 \textbar\ 0 & \xmark*  \\
Func. level rando.~\cite{conti2016selfrando} & FB & 46.5 & 43.8 & 24.2 \textbar\ 71.1 & 15.9 \textbar\ 31 & 41.2 \textbar\ 65.4 & 56.9 \textbar\ 25 & 34.5 \textbar\ 78.7 & 23 \textbar\ 75.8 & 3.5 \textbar\ 14.5 & \cmark  \\
Func.+Reg. level rando.~\cite{homescu2013profile} & FB \& Reg. & 44.2 & 43.9 & 35.5 \textbar\ 44.8 & 35.3 \textbar\ 43.4 & 63.2 \textbar\ 61.8 & 44.8 \textbar\ 49.0 & 36.4 \textbar\ 52.1 & 43.1 \textbar\ 35.3 & 66.7 \textbar\ 0 & \cmark  \\ 
Block level rand.~\cite{koo2018compiler} & BB & 20.98 & 37.0 & 7.3 \textbar\ 36.3 & 8.1 \textbar\ 32.1 & 13.9 \textbar\ 55.9 & 24.8 \textbar\ 31.6 & 22.2 \textbar\ 52.1 & 18.1 \textbar\ 44.6 & 50.0 \textbar\ 0 & \cmark  \\ \hline
\multicolumn{12}{r}{* For instruction-level randomization scheme~\cite{hawkins2017zipr}, TC is not preserved for minimum footprint gadgets, but TC is preserved for extended footprint gadgets.} \\
\end{tabular}
\end{table*}

The Turing-complete expressiveness of ROP gadgets is preserved in the randomized versions of applications or libraries when the schemes are function, block, and function+register-level randomizations. %randomized by SR, MCR, and CCR. 
%CCS20% The Turing completeness is also preserved for both minimum and extended footprint gadgets.
However, instruction-level randomization scheme~\cite{hawkins2017zipr} does not retain the Turing-complete expressiveness for minimum footprint gadgets. The Turing-complete expressiveness is hampered when there is no gadget in one of the Turing-complete operations. For example, in Table~\ref{tab:summary-main}, the reduction of minimum footprint gadgets in memory and arithmetic operations is almost 100\% for applications. That means there is no gadget to do memory and arithmetic operations, which are required for reliable attacks. The reductions for libraries in the two categories (i.e., memory and arithmetic) are 93.7\% and 91.8\%, respectively. For both application and library cases, the reductions are not exactly 100\%, because some applications/libraries contain a few gadgets. When the numbers of gadgets are averaged over the number of applications/libraries, the average is close to zero. 

Most of the applications and libraries do not contain any {\em syscall} gadgets (as expected), as applications and libraries usually make syscalls through {\em libc}. This is why the number of {\em syscall} gadgets is low (2-3) and one gadget loss leads to around 33\% reduction. Since SR is only able to randomize a light-weight version of libc ({\em musl}), we observe slightly high values (84) and low reduction for system gadgets in Table~\ref{tab:summary-main} for the function-level scheme.

%Most of the applications and libraries do not contain any {\em syscall} gadgets (as expected) because applications and libraries usually make syscalls through {\em libc}. 
%CCS20% However, applications or libraries may have occasional use of the \verb1syscall1 function. For example, the \verb1log_tid()1 function is the only one of \verb1httpd1 that invokes \verb1syscall1 function. Similarly, other applications or libraries occasionally invoke the \verb1syscall1 function. 
%This is why the number of {\em syscall} gadgets is low in most cases. %The average number of \verb1syscall1 gadgets before and after applying randomization in Table \ref{tab:tc-breakdown-min-fp} indicates unavailability of \verb1syscall1 gadgets. 
%Since SR is only able to randomize a light-weight version of libc ({\em musl}), we see slightly high values for system gadgets in Table \ref{tab:summary-main} for the function-level scheme.

We also assess the gadget availability under a {\em single} randomization pass of Shuffler~\cite{williams2016shuffler}. We take 100 consecutive address space snapshots from {\em nginx} after each re-randomization with an interval of 30 seconds and manually find gadgets from the snapshots. On average, we observe a 24\% and 3\% reduction in gadget availability for minimum and extended footprint gadgets compared to a non-randomized version of {\em nginx}, respectively. The low reductions are expected, as Shuffler's security relies on the capability of continuously shuffling code locations, not a single randomization pass.

Ideally, function-level randomization does not break gadgets, only shifts the gadgets from one location to another. Basic-block or machine-register-level randomization may break some gadgets due to the memory layout perturbation and register allocation randomization. It is not surprising that the function, block or register-level randomization schemes have low gadget reduction. On the other hand, instruction-level randomization perturbs the memory layout significantly. That is why we observe a large reduction in gadget availability by Zipr.

\noindent\textit{Future directions.} 
%We conservatively define the upper bound of a re-randomization scheme by the time required for an attacker to achieve 100\% of the Turing-complete gadget set. However, in reality, partial Turing-complete gadget set could be sufficient for an attacker in some scenarios. Thus, the relevance analysis of partial Turing-complete gadget set could be an interesting research direction.
%
%\noindent
Redefining traditional ROP gadgets into smaller (e.g., one line) building blocks and demonstrating new gadget chain compilers (e.g., two-level construction) by tackling the instruction-level perturbations are interesting new attack directions.

\color{black}
\subsection{Impact on Performance Overhead}
We measure the performance overhead of the five (re-)randomization tools to evaluate the overhead in our measurement environment. To measure the performance overhead, we use 8 applications in domains such as web servers, FTP servers, browsers, security protocols, and file compression tools. The applications are {\em nginx}, {\em httpd}, {\em proftpd}, {\em hiawatha}, {\em lighttpd}, {\em openssl}, {\em firefox}, and {\em bzip}. Applications are randomized using the five (re-)randomization tools. We use criteria such as HTTP request latency, FTP upload speed, browser page-load time, compression time, and effectiveness of cryptographic algorithms to measure the performance overhead. 

We measure HTTP request latency by running an HTTP benchmark using {\em wrk}~\cite{glozer2018wrk} for 30 seconds to read an HTML page from a server. The benchmark includes 12 threads and 400 HTTP open connections. To measure FTP upload speed, we run a benchmark using {\em ftpbench}~\cite{ftpbench2020}. The benchmark runs 10 concurrent operations for 10 seconds. We use OpenSSL {\em speed} to test the performance of aes-128-gcm, aes-256-gcm, aes-128-cbc, and aes-256-cbc algorithms. We use the Linux {\em time} command to measure compression time. Finally, we use a website speed test tool~\cite{uptrends2020} to measure a browser's page load time. For Shuffler, we measure the overhead for three different re-randomization intervals: 10ms, 100ms, and 1s. 

We run each measurement for five times and calculate the average for each application. Then, we average the overheads over the 8 applications. For Shuffler, we observe 3\% overhead with 1s re-randomization interval, 5\% for 100ms, and 12\% for the 10ms interval consistent with the reported result~\cite{williams2016shuffler}. We observe 23\% overhead for Zipr, 10\% for SR, 3\% for CCR, and 10\% for MCR which are comparable to or higher than what’s reported. The reported overheads for Zipr,  SR, CCR, and MCR are around 5\%~\cite{hawkins2017zipr}, 1\%~\cite{conti2016selfrando}, 0.28\%~\cite{koo2018compiler}, and 1\%~\cite{homescu2013profile}, respectively.
\color{black}

%----------------------------------------------------------------------------
\subsection{Impact on the Quality of a Gadget Chain} \label{evaluation-payload-generation}
%----------------------------------------------------------------------------
%CCS20% \noindent \textbf{\textit{RQ \#3}}: How does fine-grained code randomization impact the quality of a gadget chain (i.e., payload)?
The purpose of this analysis is to estimate the quality of a gadget chain. We measure the quality of a gadget through the register corruption analysis for individual gadgets, following the procedure described in Section~\ref{exp-payload-gen}. We measure the register corruption rate for \texttt{MV}, \texttt{LR}, \texttt{AM}, \texttt{LM}, \texttt{AM-LD}, \texttt{SM}, \texttt{AM-ST}, \texttt{SP}, and \texttt{CALL} gadgets.  Some gadgets such as \texttt{CP}, \texttt{RF}, and \texttt{EP} (described in Table~\ref{gadget_type} in the Appendix) are special purpose gadgets that are used to trick defense mechanisms, such as CFI~\cite{abadi2005control}, kBouncer~\cite{pappas2013transparent}, and ropecker~\cite{cheng2014ropecker}. Thus, we omit these gadgets from the quality analysis. 

We found that the overall register corruption rate is slightly higher ($\sim$6\%) in the presence of fine-grained randomization. This slightly higher register corruption rate indicates that the formation of gadget chain is slightly harder in fine-grained randomization compare to the coarse-grained randomization.

We present the detailed results in Table~\ref{register-corruption} in the Appendix, including the average number of unique registers used in each gadget. We observe the number of unique registers used in each gadget ranges from 1 to 4 in our register corruption measurement.

%CCSR2020 We present the detailed results in Appendix (Table \ref{register-corruption}). Table \ref{register-corruption} also reports the average number of unique registers used in each gadget. This number reflects how many registers (ranging from 1 to 4) are involved in a gadget on average.

Sometimes, fine-grained randomization decreases the register corruption rate. For example, for Nginx, the corruption rate of the load memory (LM) gadgets is reduced from 44\% to 15\%, when fine-grained randomization is in place. This reduction is likely due to the relatively smaller number of gadgets in the presence of the fine-grained randomization.

\noindent\textit{Future directions.} %CCS20% Most randomization solutions reorder functions, basic-blocks, and instructions. MCR~\cite{homescu2013profile} goes one step deeper, reorders machine registers, and replaces \verb|mov reg1, reg2| instructions with equivalent \verb1lea1 instruction. Besides these, designing randomization solutions that increase the register corruption rate in gadgets would be interesting as high register corruption rate would make attacks unreliable. 
%MCR~\cite{homescu2013profile, crane2015readactor} reorders machine registers and replaces {\em mov reg1, reg2} instructions with equivalent {\em lea} instruction. 
Designing randomization solutions to increase the register corruption rate in gadgets would be interesting as a high register corruption rate would make attacks unreliable. 

\subsection{Availability of Libc Pointers} \label{evaluation-libc-pointers}
%------------------------------------------------------------------
%\textbf{\textit{RQ \#3}}: How useful is a heap or stack or data segment memory leak for revealing the libc library (i.e., revealing libc functions)?

\noindent This experiment measures the risks associated with a heap, stack or data segment of an application for revealing a library location. For simplicity, we consider only the risk associated with revealing the {\em libc} library w.r.t. the basic ROP attacks. We count the number of unique {\em libc} pointers in a target application's stack, heap, and data segment when the application reaches a certain execution point. The execution point is defined differently for different types of applications. For example, the execution point for {\em proftpd} is when {\em proftpd} is ready to accept connections. We assume that {\em i)} coarse-grained randomization is enforced, and {\em ii)} adversaries are not able to perform recursive code harvest to find gadgets. This experiment targets a weak attack model where an adversary leaks a (known) library pointer and adjusts pre-computed gadgets based on the leaked pointer. We regard a library pointer (e.g., {\em libc} pointer) as known if the pointer is loaded in the same location in the stack of an application for multiple runs.
A pointer in stack, heap or data segment may point to a non-library function, which in turn points to a library (e.g., {\em libc}). %Recursive code harvesting can then utilize such a pointer to access libraries.

\begin{figure}[!htp]
    \centering
    \includegraphics[width=0.40\textwidth]{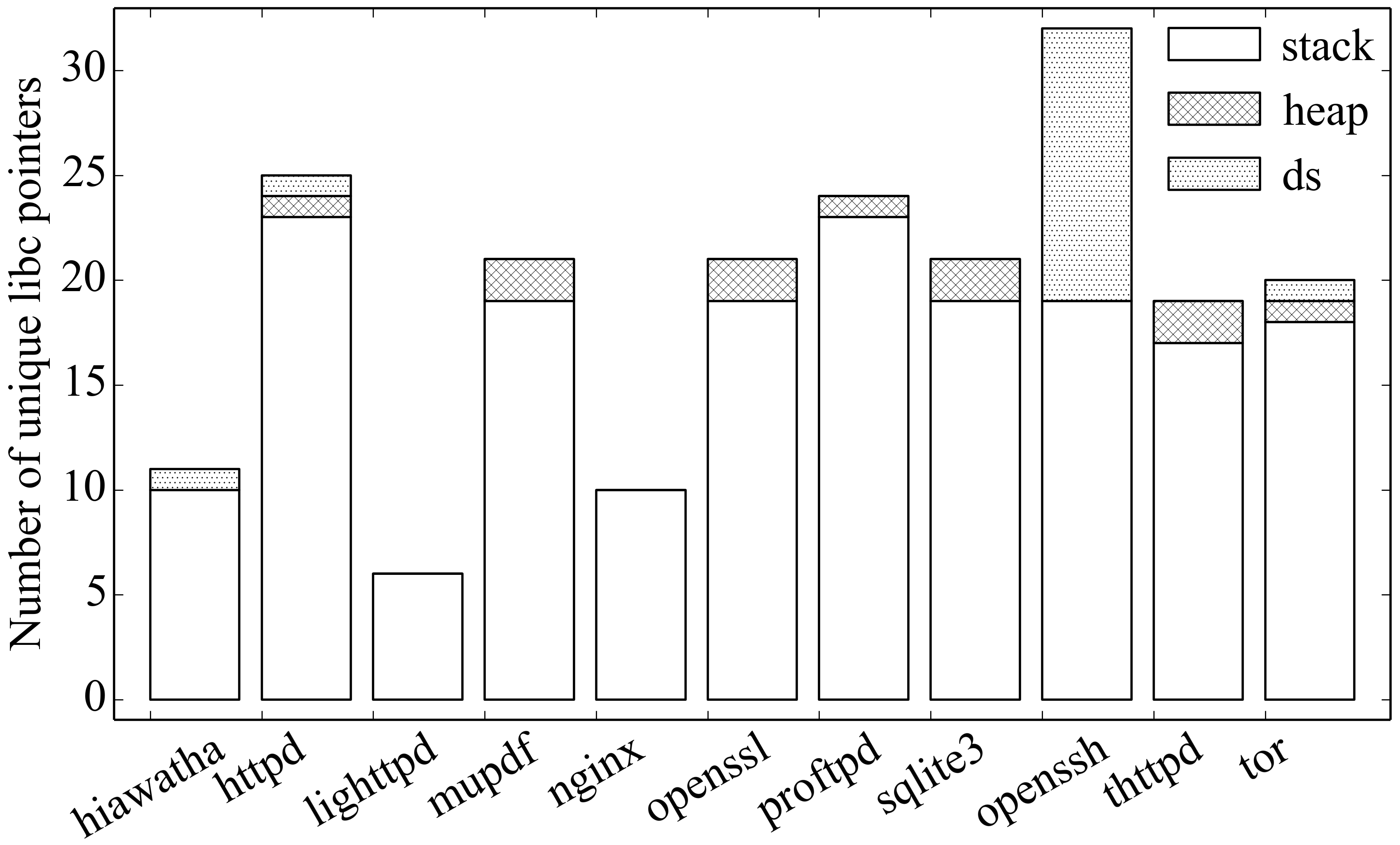}
    \caption{Libc pointers in the stack, heap and data segment of a program. Stacks contain more pointers, carrying higher risks of pointer leakage.}
    %\caption{Times needed to harvest the Turing-complete gadget set. The minimum, maximum, and average times are calculated from the times that the code harvest process takes by starting from a code pointer leaked from each code page.}
    \label{fig:libc-pointers-hs}
\end{figure}

Figure~\ref{fig:libc-pointers-hs} shows the number of unique {\em libc} pointers %and the number of times those pointers appear 
in the stack, heap, and data segment of 11 applications including web servers, PDF reader, cryptography library, database, and browser. According to the observations in Figure~\ref{fig:libc-pointers-hs}, heap or data-segment contains only one {\em libc} code pointer (on average) while stack contains 17 {\em libc} code pointers. This finding indicates that high risk is associated with stack than heap or data segment. It also suggests that the safeguard and randomization/re-randomization of stack is more important than protecting/randomizing heap or global variables. 

%For most of the programs in Figure~\ref{fig:libc-pointers-hs}, none of the heaps or data segments contains more than three libc pointers (pointing to libc's text segment). In contrast, stacks do have many libc pointers. Though it is tough to leak arbitrary pointers from stack, heap or data segment in practice, a high risk is associated with stack than heap or data segment.

%\noindent\textbf{Summary}: A stack has a higher risk of revealing dynamic libraries than a heap or data segment.

\noindent\textit{Future directions.} %Three (i.e., Zipr, SR, and CCR) out of the four randomization solutions are designed to randomize only code so that information leakage becomes useless. 
%CCS20% MCR protects stack by inserting padding, randomizing stack frames, and randomizing the locations of global variables. These protections are important as they stop memory leakage or harden memory leakage processes in the first place. Thus, protecting a stack/heap/data-segment from leaking code pointers along with data pointers that contain code pointers would be an interesting research direction.
Protecting a stack/heap/data-segment from leaking data pointers that contain code pointers would be an interesting research direction. \color{black}A similar risk assessment for C++ binaries could indicate the importance of protecting the read-only section that includes many pointers to virtual methods.\color{black}

% uncomment for lists of compiler optimizations :END

\section{Discussion} \label{sec:discussion}

\noindent
{\textbf{Metrics for evaluating fine-grained randomization}}. Traditionally, both coarse- and fine-grained randomization solutions use entropy to measure the effectiveness of hindering code-reuse attacks~\cite{team2003pax, conti2016selfrando, koo2018compiler, wartell2012binary}. 
%{\bf Salman, please add references here. I remember ccr or selfrando used entropy, see if there're others.} 
%
%However, the entropy metric alone may not be enough for assessing the prevention of code-reuse attacks given that attackers have JIT-ROP capabilities. It is not enough because attack elements may remain in a program randomized with a good entropy value.
%
Randomization tools such as PaX ASLR~\cite{team2003pax},  SR~\cite{conti2016selfrando}, CCR~\cite{koo2018compiler}, Remix~\cite{chen2016remix}, Binary stirring~\cite{wartell2012binary}, ILR~\cite{hiser2012ilr} and ASLP~\cite{kil2006address} use the entropy values as the security metrics to evaluate the security of their randomization schemes. Some tools such as SR~\cite{conti2016selfrando}, CCR~\cite{koo2018compiler}, Remix~\cite{chen2016remix}, and ASLP~\cite{kil2006address} calculate entropy as a function of the number of function- or basic-blocks. That means these tools permute the order of functions or basic blocks.

%CCS20% Some tools such as SR~\cite{conti2016selfrando}, CCR~\cite{koo2018compiler}, Remix~\cite{chen2016remix}, and ASLP~\cite{kil2006address} calculate the entropy value as a function of the number of functions and basic blocks. That means these tools permute the order of functions or basic blocks of program code and place it different places in the address of that program. These tools also consider some constraints such as executable size and fall-through basic blocks. 

However, such an entropy measure is not useful under the JIT-ROP threat model, as chunks of code are still available. Including distances between permuted functions or basic blocks in the entropy computation would not work either, because the code's semantic connectivity (e.g., through {\em call} and {\em jmp}) is still not captured. Code connectivity is what JIT-ROP attacks leverage to discover code pages. In comparison, our measurement methodology more accurately reflects JIT-ROP capabilities and is more meaningful under the JIT-ROP model. How to design an entropy-like metric to capture the degree of code isolation or the {\bf semantic connectivity} in code is an interesting open problem.

\noindent
\label{availability-bop}{\textbf{Availability of Block-Oriented Programming (BOP) gadgets}}.
We measure the numbers of BOP functional blocks for register assignments/modifications, memory reads/writes, system/library calls, and conditional jumps using the BOP compiler (BOPC)~\cite{ispoglou2018block}. We observed almost no change in the numbers of BOP functional blocks in randomized versions compared to the non-randomized versions for CCR~\cite{koo2018compiler} and MCR~\cite{homescu2013profile, crane2015readactor}.  As BOPC operates on static binary, we could not use Shuffler~\cite{williams2016shuffler} and SR~\cite{conti2016selfrando} because they randomize a memory layout at runtime. BOPC does not seem to run on binaries produced by Zipr~\cite{hawkins2017zipr}.

\noindent
{\bf Impact of the compiler optimizations on gadget availability}. We assess the impact of code transformations and optimizations (-O0, -O1, -O2, -O3, -Ofast, -Os) on the availability of gadgets. We compare the unoptimized and optimized versions of {\em nginx}, {\em apache}, {\em proftpd}, {\em openssh}, and {\em sqlite3} to assess the impact. We find that the unoptimized code contains a smaller number of LM, SM, and MR gadgets than optimized code (Figure~\ref{fig-optim-plot} in the Appendix). The main reason is due to the tendency of {\em mov} and {\em ret} instructions staying together in optimized code, but not in unoptimized code. Besides, compilers sometimes emit extra instructions for optimizations that increase gadgets.

\noindent
{\textbf{Reachability of gadgets}}. We design our experiments based on the availability of various kinds of gadgets. However, in reality, it is not an easy task to invoke the available gadgets. Attackers need to conduct a series of operations including finding a vulnerability or leaking memory for the actual invocations of gadgets.  In Section~\ref{threat-model-assumptions}, we assume that an attacker has already overcome the initial obstacles, especially finding a memory leak. Our experiments are focused on the available gadgets utilizing the leaked memory to compare various code (re-)randomization techniques.

\noindent
{\textbf{Operational re-randomization intervals}}. Our methodology helps guide software owners (e.g., server owners) to set the appropriate re-randomization intervals. For example, if the owners prioritize performance over security, they can set an interval just below $\mathcal{T}_\mathcal{P}^\mathcal{A}$ (Definition~\ref{upperbound}). If the owners prioritize security over performance, they can set an interval much shorter than $\mathcal{T}_\mathcal{P}^\mathcal{A}$.

%CCSR2020 Our methodology helps guide software owners (e.g., server owners) to set the appropriate re-randomization intervals. For example, if the owners prioritize performance over security, they can set an interval as the time just before when the gadget discovery process converges to a gadget set. If the owners prioritize security over performance, they can consider setting the interval as the time when the gadget discovery process achieves around 60--70\% gadgets from the gadget set.

%Daphne moved this paragraph here based on Fabian's complaint
%---------------------------------------------------------------------------------
\noindent
{\textbf{Need for randomizing Glibc}}. \label{glibc-diversification}
%---------------------------------------------------------------------------------
Unfortunately, SR, CCR, MCR, and Zipr were all unable to completely randomize the Glibc implementation. For CCR and MCR, the LLVM Clang compiler (which CCR and MCR use as their compilers) does not have the support for certain GCC specific extensions (e.g., ASM GOTO) in Glibc. 
%Glibc contains GCC specific non-standard extensions (e.g., ASM GOTO), which Clang does not cover. 
SR also cannot randomize some parts of Glibc. Therefore, we evaluate a lightweight version of the standard C library \verb1musl-libc1~\cite{musl-libc}) instead of Glibc. Only Selfrando works on \verb1musl-libc1. %However, we are unable to diversify it with the other tools. On the other hand,
Shuffler can reorder Glibc code by making a few modifications such as disabling manual jump table construction.

\noindent
{\textbf{Limitations.}}
\color{black}
Both CFI and XoM defenses are powerful and have capabilities to prevent JIT-ROP attacks. These two defenses with continuous re-randomization would be even more powerful. However, we did not enforce CFI and XoM in this work so that we could isolate an individual defense’s security impact. In this work, we attempt to address many important questions related to fine-grained (re-)randomization not yet answered by the literature. We leave the analysis and measurement of CFI and XoM as a future research direction.
\color{black}
%Fine-grained code randomization defenses \cite{conti2016selfrando, hiser2012ilr, homescu2013profile, larsen2014security, koo2018compiler, pappas2012smashing} can protect programs from traditional ROP attacks. Unfortunately, these defenses are vulnerable to information leak \cite{serna2012info} attacks. Snow et al. demonstrated this vulnerability of fine-grained randomization solutions using JIT-ROP attack \cite{snow2013just}. 

Our current work does not measure zombie gadgets~\cite{snow2016return} and microgadgets~\cite{homescu2012microgadgets}.
The gadgets that are available after applying destructive read defenses (e.g., XnR~\cite{backes2014you}, NEAR~\cite{werner2016no}, Readactor~\cite{crane2015readactor}, and Heisenbyte~\cite{tang2015heisenbyte}) are called {\em zombie gadgets}~\cite{snow2016return}. Destructive read defenses only allow code execution, no read after execution. %Any attempt to read code pages terminate a process. 
In this way, destructive reads destroy the availability of gadgets to attackers. However, destructive read defenses cannot eliminate all gadgets.
In our future work, we plan to assess the availability of zombie and micro gadgets after applying destructive read defenses. %CCSR2020% In particular, we plan to assess the entity (i.e., JIT compilers or load/unload feature or new process creation or implicit reads) that can facilitate attacks by providing many gadgets. %CCS20% We will categorize the available zombie gadgets in seven Turing-complete (TC) categories and measure the availability of zombie gadgets in different TC categories. 
%We will perform a large-scale evaluation on the availability of zombie gadgets for various applications including popular web servers, PDF readers, browsers, databases, and crypto libraries. 
%We plan to assess the feasibility of implicit reads \cite{snow2016return} after applying various code transformations such as instruction substitution, basic block instruction reordering, register preservation code reordering, register reassignment, and instruction displacement through fine-grained code randomization and re-randomization.

Another limitation is that our threat model assumes that code pointer obfuscation-based defense is not deployed. If used, code pointer obfuscation (e.g., CPI~\cite{kuznetsov2014code, kuznetzov2018code}, Oxymoron~\cite{backes2014oxymoron}) could make JIT-ROP code page discovery less effective, reducing the gadget availability. Understanding how code pointer obfuscation impacts JIT-ROP and measuring the effectiveness of these defenses under various attack conditions (e.g.,  Isomeron~\cite{davi2015isomeron} and COOP~\cite{schuster2015counterfeit}) are interesting problems.

%CCSR2020 Another limitation is that our threat model assumes that code pointer obfuscation-based defense is not deployed. If used, code pointer obfuscation could make JIT-ROP code page discovery less effective, reducing the gadget availability. For example, Oxymoron~\cite{backes2014oxymoron} showed some effectiveness for obstructing JIT-ROP by making code pointer redirection through a randomization-agnostic translation table. CPI proposed by Kuznetsov {\it et al.}~\cite{kuznetsov2014code} is shown to be effective for mitigating JIT-ROP and COOP attacks. Understanding how code pointer obfuscation impacts JIT-ROP and measuring the effectiveness of these defenses under various attack conditions (e.g.,  Isomeron~\cite{davi2015isomeron} and COOP~\cite{schuster2015counterfeit}) are interesting problems.

\color{black}
One limitation of the time-based re-randomization schemes is that the time needs recalculation with the evolution of hardware or a program itself. Event-based re-randomization schemes can be effective in this case. However, event-based schemes may trigger unnecessary re-randomization if events are frequent, e.g., re-randomizing every time a program outputs~\cite{bigelow2015timely}.
\color{black}

\vspace{5pt}

\noindent
{\bf Key Takeaways}\label{sec:key-takeaways}
%-------------------------------------------------------------------------

%\begin{enumerate}

    % \item 
    %RQ1
    %commented this for AsiaCCS
    %\noindent{\em Fine-grained code randomization up to basic block level does not substantially weaken attackers' capabilities, however, instruction-level does.}
    %commented this for AsiaCCS
    %Compared with coarse-grained code randomization, fine-grained randomization reduces the availability of minimum footprint gadgets by up to \textbf{40\%} and extended footprint gadgets by up to \textbf{38\%} when the function-level, basic-block-level, or machine register-level randomization is applied. This reduces the availability of gadgets but does not eliminate the possibility of ROP attacks. However, instruction-level fine-grained randomization eliminates the possibility of ROP attacks.
   %This observation somewhat contradicts the intuition that function level randomization would not impact the total number of gadgets. We explain the possible causes (e.g.,  \verb1ret1 structures got changed).
    
    %CCSR2020\noindent {\em Effective re-randomization upper bound}. A methodology for measuring the Turing-complete gadget set systematically by considering the speed of gadget convergence can help compute the effective upper bound for re-randomization intervals of a re-randomization scheme. \color{blue}Our experiments show that this upper bound ranges from around 0.8 to 4.3 seconds for various applications/browsers.\color{black} Applying our methodology on their machines will help re-randomization adopters to make more informed configuration decisions.
    
    \noindent \ding{182} {\em Effective re-randomization upper bound}. Our methodology for measuring various gadget sets systematically by considering the gadget convergence time helps compute the effective upper bound for re-randomization intervals of a re-randomization scheme. \color{black}Our results show that this upper bound ranges from around 1.5 to 3.5 seconds. \color{black} Applying our methodology on their machines will help re-randomization adopters to make more informed configuration decisions.

   \noindent \ding{183} {\em All leaked pointers are created equal for gadget convergence, but not for the speed of gadget convergence}. Regardless of the location of pointer leakage, we obtain the same number of minimum and extended footprint gadgets via JIT-ROP. This observation indicates that any pointer leak from an application's code segment is equally useful for attackers. However, the time for obtaining the gadgets varies for different leaked pointers.
    
    %CCS20% \noindent {\em All leaked pointers are created equal}.
    %CCS20% Regardless of the location of pointer leakage, we are able to obtain the same number of minimum and extended footprint gadgets via JIT-ROP. This observation indicates that any pointer leak from an application's code segment is equally useful for attackers. Any leaked pointers would enable attackers to harvest a large number of code pages and gadgets.
    %The location of the leaked pointer has only a small impact on the number of discovered gadgets. Regardless of whether PIE is turned on or not, a pointer leakage in any location allows attackers to access a set of MIN-FP gadgets. 
    
    %Commented out based on Fabian's suggestion
   % \noindent {\em Future research directions.}
    %Our findings suggest a range of exciting new attack and defense directions, including redefining traditional gadget finder and chain compilers, evaluating fine-grained randomization solutions based on code isolation and register corruption rate metrics, and safeguarding and randomizing stack.
    
   \noindent \ding{184} {\em  Turing-complete operations}. Function, basic-block, or machine register level fine-grained randomization preserves Turing-complete expressive power of ROP gadgets, however, instruction-level randomization does not. %Besides, unoptimized code seems to limit more Turing-complete operations than optimized code.
    
    \noindent \ding{185} {\em Connectivity}. Code connectivity is the main enabler of JIT-ROP. As the conventional entropy metric does not capture code connectivity, it should not be used to measure ASLR security under the JIT-ROP threat model. %Approaches for obfuscating code connectivity are promising in building JIT-ROP defenses.
    
    \noindent \ding{186} {\em Gadget quality}. Our findings suggest that current fine-grained randomizations do not impose significant gadget corruption. 
    
    %{\bf Salman, maybe add 1 sentence about gadget corruption somewhere in the takeaways? thanks.}
%\end{enumerate}

%-----------------------------------------------------
%\section{Notes for us}
%-----------------------------------------------------
%\begin{enumerate}
%    \item Why ccr and selfrando differs?
%    \item What is a good benchmark for evaluating a code diversification technique?
 %   \item Why diversification tools fail to diversify glibc implementation?
%    \item What is a good technique to find gadgets in diversified code? Is the traditional technique enough?
%    \item jmp, call count on diversified and undiversified, difference between ccr and selfrando
%\end{enumerate}
%-----------------------------------------------------

\section{Related Work}\label{related-work}
%-----------------------------------------

The research conducted in the system security area primarily has two  themes: 1) demonstrating attacks and 2) discovering countermeasures.  Attack demonstrations range from stack smashing~\cite{one06phrack}, return-to-libc~\cite{designer1997return, krahmer2005x86, wojtczuk2001advanced}, to ROP~\cite{checkoway2010return, kayaalp2012branch, carlini2014rop}, JOP~\cite{bletsch2011jump}, DOP~\cite{hu2016data}, ASLR bypasses~\cite{snow2013just, davi2015isomeron, bittau2014hacking, gawlik2016enabling, hu2016data, goktacs2016undermining}, and CFI bypasses~\cite{carlini2015control, goktas2014out, carlini2014rop, biondo2018back, ispoglou2018block}. 

Researchers have also proposed a range of defenses for ROP attacks~\cite {cheng2014ropecker, pappas2013transparent, davi2011ropdefender, chen2009drop, davi2009dynamic, pappas2012smashing, onarlioglu2010g, goktacs2014size, schuster2014evaluating, fratric2012ropguard, crane2015readactor, abadi2005control, zhang2013practical, zhang2013control, niu2014modular, bletsch2011mitigating, criswell2014kcofi, erlingsson2006xfi, payer2015fine, van2017dynamics}, CFI bypass~\cite{zhang2013practical}, and ASLR bypass~\cite{davi2015isomeron, lu2015aslr, backes2014oxymoron, maisuradze2016cannot, kil2006address,chen2016remix,wartell2012binary,giuffrida2012enhanced,hiser2012ilr,pappas2012smashing,koo2018compiler,backes2014you,werner2016no,crane2015readactor,tang2015heisenbyte,bigelow2015timely,williams2016shuffler}. A categorical representation of these defenses is given in our attack-path diagram (Figure~\ref{attac-decsision} in the Appendix). Binary analysis tools are also available to understand~\cite{shoshitaishvili2016sok} and mitigate~\cite{van2016tough} these ROP or code-reuse attacks.

Most of the above-mentioned defenses are variants of W$\oplus$X (e.g., NEAR~\cite{werner2016no} and Heisenbyte~\cite{tang2015heisenbyte}), memory safety (e.g., HardScope~\cite{nyman2019hardscope}, Memcheck~\cite{nethercote2007valgrind}, AddressSanitizer~\cite{serebryany2012addresssanitizer}, and StackArmor~\cite{chen2015stackarmor}), ASLR (e.g., fine-grained randomization~\cite{chen2016remix, koo2018compiler, wartell2012binary, bigelow2015timely, williams2016shuffler, seo2017sgx}), and CFI (e.g., CCFIR~\cite{zhang2013practical} and bin-CFI~\cite{zhang2013control}). These defenses are capable of preventing most code-reuse attacks~\cite{snow2013just, bittau2014hacking, davi2015isomeron, gawlik2016enabling} except a few cases such as inference attacks that are performed using zombie gadgets~\cite{snow2016return} or relative address space layout~\cite{rudd2017address, goktas2018position}. 
The latest advancement in control-flow transfers such as MLTA~\cite{lu2019does} significantly advances CFI that can prevent most control-oriented attacks. 
Recent attention on non-control-oriented or data-only attacks~\cite{hu2016data, ispoglou2018block} motivated researchers to develop practical Data-Flow-Integrity (DFI)~\cite{castro2006securing} solutions (details of non-control attacks in~\cite{cheng2019exploitation}). Currently, it is challenging to implement a practical DFI solution considering the overhead of data-flow tracking.

% ASLR, CFI, CPI, XoM, and XnR 
From the defense-in-depth perspective, it is desirable to have some degree of redundancy (e.g., CFI and ASLR) in system protection. A single deployed defense may be compromised due to unknown implementation flaws or configuration issues. Thus, investigations in multiple directions~\cite{shacham2004effectiveness, burow2017control, van2017dynamics, homescu2012microgadgets} is necessary for gauging the feasibility of existing defenses. Our work investigates various aspects of ASLR -- including timing -- by evaluating security metrics such as various gadget sets, interval choices, and code pointer leakages. We also assess how security tools in the ASLR domain impact on these security metrics, quantitatively.

\section{Conclusions} \label{sec:conclusions}
%---------------------------------------------------
We presented multiple general methodologies for quantitatively measuring the ASLR security under the JIT-ROP threat model and conducted a comprehensive measurement study. One method is for computing the number of various types of gadgets and their quality. Another method is for experimentally determining the upper bound of re-randomization intervals. The upper bound helps guide re-randomization adopters to make more informed configuration decisions. 
\section*{Acknowledgment}
%-------------------------------------------------------------------------------
We thank our shepherd, Georgios Portokalidis, for his support and valuable feedback for this work. We also thank the anonymous reviewers for their valuable comments and suggestions. 

%This work is supported by DARPA/ONR Grant N66001-17-C-4052.

%This work has been in part supported by the Office of Naval Research under Grant ONR-N00014-17-1-2498, DARPA/ONR Grant N66001-17-C-4052, National Science Foundation under Grants OAC-1541105 and CNS-1801534, Intel Collaborative Research Institute for Collaborative Autonomous \& Resilient Systems (ICRI-CARS), and the Academy of Finland under Grant 309994 (SELIoT). The views and conclusions contained herein are those of the authors and should not be interpreted as necessarily representing the official policies or endorsements, either expressed or implied, of any of the above organizations or any person connected with them.

%Any opinions, findings, conclusions, or recommendations expressed in this material are those of the authors and do not necessarily reflect the views of their employers or the sponsors.

%\section{Acknowledgments}
%\begin{verbatim}
%  \begin{acks}
%  ...
%  \end{acks}
%\end{verbatim}

%\begin{acks}
%To Robert, for the bagels and explaining CMYK and color spaces.
%\end{acks}

\bibliographystyle{ACM-Reference-Format}
\bibliography{ref}

\appendix
%---------------------------------
\section{Appendix}\label{appendix}
%---------------------------------

%==================== Gadget types table ==========================
\begin{table*}[!ht]
%\scriptsize
\footnotesize
\caption{Gadgets used in advanced ROP attacks~\cite{snow2013just,carlini2014rop, goktas2014out, bittau2014hacking, carlini2015control} . $\bigtriangleup$ indicates an addition/subtraction/multiply/division. $\upphi$ indicates logical operations such as and, or, left-shift, and right-shift. $\bigtriangledown$ indicates any operation that modifies stack pointer (SP). SN $\rightarrow$ Short name. TC? indicates whether a gadget is included in the Turing-complete gadget set or not.} %JIT-ROP $\rightarrow$ Just-In-Time Code Reuse \cite{snow2013just} CBD $\rightarrow$ Control Flow Bending \cite{carlini2015control}, EHH $\rightarrow$ Evasive and History Hiding \cite{carlini2014rop}, OOC $\rightarrow$ Out of Control \cite{goktas2014out} , and HB $\rightarrow$ Hacking Blind \cite{bittau2014hacking}.}
\label{gadget_type}
\centering
\renewcommand{\arraystretch}{1.1}
\setlength\tabcolsep{4.8pt} % default value: 6pt
\begin{tabular}{lllllll}
\hline
\textbf{Gadget types} & \textbf{Purpose} & \textbf{Minimum footprint} & \textbf{Example} & \textbf{TC?} & \begin{tabular}[c]{@{}l@{}}\textbf{SN}\end{tabular} & \textbf{Source} \\ \hline
\rowcolor{new-light-gray}
Move register & Sets the value of one register by another & mov reg1, reg2; ret & mov rdi, rax; ret & \checkmark & MR &~\cite{snow2013just} \\ %\hline
Load register & Loads a constant value to a register & pop reg; ret & pop rbx; ret & \checkmark  & LR &~\cite{snow2013just, carlini2015control} \\ %\hline
\rowcolor{new-light-gray}
Arithmetic & \begin{tabular}[c]{@{}l@{}}Stores an arithmetic operation's result of\\two register values to the first\end{tabular} & $\bigtriangleup$ reg1, reg2; ret & add rcx, rbx; ret & \checkmark  & AM &~\cite{snow2013just} \\ %\hline
Load memory & Loads a memory content to a register & mov reg1, [reg2]; ret & mov rax, [rdx]; ret & \checkmark  & LM &~\cite{snow2013just, carlini2015control} \\ %\hline
\rowcolor{new-light-gray}
Arithmetic load & \begin{tabular}[c]{@{}l@{}}$\bigtriangleup$ a memory content to/from/by a\\register and store in that register\end{tabular} & $\bigtriangleup$ reg1, [reg2]; ret & add rsi, [rbp]; ret & \checkmark  & AM-LD &~\cite{snow2013just} \\ %\hline
Store memory & Stores the value of a register in memory & mov [reg1], reg2; ret & mov [rdi], rax; ret & \checkmark & SM &~\cite{snow2013just} \\ %\hline
\rowcolor{new-light-gray}
Arithmetic store & \begin{tabular}[c]{@{}l@{}}$\bigtriangleup$ a register value to/from/by a memory\\content and stores in that memory\end{tabular} & $\bigtriangleup$ [reg1], reg2; ret & sub [ebx], eax; ret & \checkmark & AM-ST &~\cite{snow2013just} \\ %\hline
Logical & \begin{tabular}[c]{@{}l@{}}Performs logical operations\end{tabular} & \begin{tabular}[c]{@{}l@{}} $\upphi$ reg1, reg2; ret \\ $\upphi$ reg1, const; ret \\ $\upphi$ [reg1], reg2; ret \\ $\upphi$ [reg1], const; ret \end{tabular} & shl rax, cl; ret; & \checkmark & LOGIC &~\cite{roemer2012return} \\ %\hline
\rowcolor{new-light-gray}
Stack pivot & Sets the stack pointer, SP & $\bigtriangledown$ sp, reg & xchg rsp, rax & $\times$ & SP &~\cite{snow2013just} \\ %\hline
Jump & Sets instruction pointer, EIP. & jmp reg & jmp rdi & \checkmark & JMP &~\cite{snow2013just} \\ %\hline
\rowcolor{new-light-gray}
Call & \begin{tabular}[c]{@{}l@{}}Jumps to a function through a register\\or memory indirect call\end{tabular} & call reg or call [reg] & call rdi & \checkmark & CALL &~\cite{snow2013just} \\ %\hline
System Call & \begin{tabular}[c]{@{}l@{}}Invokes system functions\end{tabular} & syscall or int 0x80; ret & syscall & \checkmark & SYS &~\cite{roemer2012return} \\ %\hline
\rowcolor{new-light-gray}
Call preceded & Bypasses call-ret ROP defense policy & \begin{tabular}[c]{@{}l@{}}mov [reg1], reg2;\\call reg3\end{tabular} & \begin{tabular}[c]{@{}l@{}}mov [rsp], rsi;\\call rdi\end{tabular} & $\times$ & CP &~\cite{carlini2015control} \\ %\hline
Context switch & \begin{tabular}[c]{@{}l@{}}Allows processes to write to Last \\ Branch Record (LBR) to flash it\end{tabular} & long loop. & \begin{tabular}[c]{@{}l@{}}3dd4: dec, ecx\\ 3dd5: fmul, [BC8h]\\ 3ddb: jne, 3dd4\end{tabular} & $\times$ & CS1 &~\cite{carlini2015control} \\ %\hline
\rowcolor{new-light-gray}
Flashing & \begin{tabular}[c]{@{}l@{}}Clears the history of LBR\\(Last Branch Record)\end{tabular} & \begin{tabular}[c]{@{}l@{}}Any simple call\\ preceded gadgets with\\a ret instruction\end{tabular} & \begin{tabular}[c]{@{}l@{}}jmp A\\ ... \\ A: mov rax, 3; ret;\end{tabular} & $\times$ & FS &~\cite{carlini2014rop} \\ %\hline

Terminal & \begin{tabular}[c]{@{}l@{}}Bypasses kBouncer heuristics\end{tabular} & \begin{tabular}[c]{@{}l@{}}Any gadgets that are \\20 instructions long\end{tabular} & \begin{tabular}[c]{@{}l@{}}N/A\end{tabular} & $\times$ & TM &~\cite{carlini2014rop} \\ %\hline
\rowcolor{new-light-gray}
Reflector & \begin{tabular}[c]{@{}l@{}}Allows to jump to both call-preceded\\or non-call-preceded gadgets\end{tabular} & \begin{tabular}[c]{@{}l@{}}mov [reg1], reg2; \\ call reg3; ... ; jmp reg4\end{tabular} & \begin{tabular}[c]{@{}l@{}}mov [rsp], rsi; \\ call rdi; ... ; jmp rax\end{tabular} & $\times$ & RF &~\cite{carlini2015control} \\ %\hline

Call site & \begin{tabular}[c]{@{}l@{}}This gadget chains the control to go\\ forward when we have the control\\on the stack and ret\end{tabular} & \begin{tabular}[c]{@{}l@{}}call reg or call [reg];\\ ...\\ ret;\end{tabular} & \begin{tabular}[c]{@{}l@{}}call rdi;\\ ...\\ ret;\end{tabular} & $\times$ & CS2 &~\cite{goktas2014out} \\ %\hline
\rowcolor{new-light-gray}
Entry point & \begin{tabular}[c]{@{}l@{}}This gadget chains the control to go\\ forward when we have the control\\of a call instruction\end{tabular} & \begin{tabular}[c]{@{}l@{}}pop rbp;\\ ...\\ call/jmp reg or\\call/jmp [reg]\end{tabular} & \begin{tabular}[c]{@{}l@{}}pop rbp\\ ...\\ call/jmp reg or\\ call/jmp [reg]\end{tabular} & $\times$ & EP &~\cite{goktas2014out} \\ %\hline

BROP & Restores all saved registers & \begin{tabular}[c]{@{}l@{}}pop rbx; pop rbp;\\pop r12; pop r13;\\pop r14; pop rsi;\\ pop r15; pop rdi;\\ret;\end{tabular} & \begin{tabular}[c]{@{}l@{}}pop rbx; pop rbp;\\pop r12; pop r13;\\ pop r14; pop rsi; \\ pop r15; pop rdi;\\ret;\end{tabular} & $\times$ & BROP &~\cite{bittau2014hacking} \\ %\hline
\rowcolor{new-light-gray}
Stop & Halts the program execution & Infinite loop & \begin{tabular}[c]{@{}l@{}}4a833dd4: inc rax\\ 3ddb: jmp 3dd4\end{tabular} & $\times$ & STOP &~\cite{bittau2014hacking} \\ %\hline
\end{tabular}
\end{table*}
%=============================================================================

%%%%%%%%%%%%%%% MOV TC and Priority Gadget set %%%%%%%%%%%%%%%%%%%%%%%%%%%
\begin{table}[!h]
\footnotesize
\caption{Gadgets with gadget types in the priority and MOV TC gadget sets.}
\label{tab:gadgets-priority-movtc-sets}
\begin{tabular}{ll|ll}
 & Priority & \textbf{} & MOV TC \\ \hline
\textbf{Type} & \textbf{Gadget} & \textbf{Type} & \textbf{Gadget} \\ \hline
\rowcolor{new-light-gray}
LR & \begin{tabular}[c]{@{}l@{}}1. pop reg\\ 2. pop reg; pop reg\end{tabular} & MR & 1. mov reg, reg/const \\
AM & 3. add reg, const & ST & 2. mov {[}reg{]}, reg \\
\rowcolor{new-light-gray}
LM & 4. mov reg, {[}reg{]}; ret & STCONSTEX & 3. mov {[}reg+offset{]}, reg/const \\
\rowcolor{new-light-gray}
JMP & 5. jmp reg & STCONST & 4. mov {[}reg{]}, const \\
ST & 6. mov {[}reg{]}, reg; ret & LM & 5. mov reg, {[}reg{]} \\
\rowcolor{new-light-gray}
SP & 7. xchg rsp, reg & LMEX & 6. mov reg, {[}reg+offset{]} \\
LOGIC & \begin{tabular}[c]{@{}l@{}}8. xor reg, reg\\ 9. xor reg, const\end{tabular} & SYS & 7. syscall \\
\rowcolor{new-light-gray}
MR & \begin{tabular}[c]{@{}l@{}}10. mov reg, reg\\ 11. mov reg, const\end{tabular} &  &  \\
CALL & \begin{tabular}[c]{@{}l@{}}12. call reg\\ 13. mov reg, reg, call reg\end{tabular} &  &  \\
\rowcolor{new-light-gray}
SYS & 14. syscall &  & 
\end{tabular}
\end{table}

%%%%%%%%%%%%%%%%%%%%%%%%%%%%%%%%%%%%%%%%%%%%%%%%%%%%%%%%%%%%%%%%%%%%

%==================== Key differences between randomization tools ==================

\begin{table*}[!ht]
\centering
\scriptsize
%\footnotesize
%\renewcommand{\arraystretch}{1.3}
\setlength\tabcolsep{3pt} % default value: 6pt
\caption{Key differences in various randomization and re-randomization schemes evaluated.}
\label{tab:summary-rand-tools}
\begin{tabular}{llllll}
Tools & \begin{tabular}[c]{@{}l@{}}Randomization\\Scheme(s)\end{tabular} & \begin{tabular}[c]{@{}l@{}}Randomization\\Time\end{tabular} & \begin{tabular}[c]{@{}l@{}}Compiler\\Assistance\\ Required?\end{tabular} & Techniques & \begin{tabular}[c]{@{}l@{}}Performance\\Overhead\end{tabular} \\
\hline
Shuffler~\cite{williams2016shuffler} & \begin{tabular}[c]{@{}l@{}} Function-level\\ re-randomization\end{tabular} & Runtime & No & \begin{tabular}[c]{@{}l@{}}- Loads itself as a user space program\\ - Contains a separate thread for shuffling the functions continuously\end{tabular} & 14.9\%~\cite{williams2016shuffler} \\
%Shuffler~\cite{williams2016shuffler} & \begin{tabular}[c]{@{}l@{}} Function-level\\ re-randomization\end{tabular} & Runtime & No & \begin{tabular}[c]{@{}l@{}}- Loads itself as a user space program\\ - Contains a separate thread for shuffling the functions continuously\\- Represents code pointers as indices for flexibility\end{tabular} & 14.9\%~\cite{williams2016shuffler} \\
\hline
Zipr~\cite{hawkins2017zipr} & \begin{tabular}[c]{@{}l@{}}Instruction-level\\ randomization\end{tabular} & \begin{tabular}[c]{@{}l@{}}Static\\rewriting\end{tabular} & No & \begin{tabular}[c]{@{}l@{}}- Reorders all instructions and generates ILR static rewrite rules\\ - Executes randomly scatter instructions using a process-level virtual\\ \hspace{0.155cm}machine (PVM) utilizing static rewrite rules or a fall-through map\\ - Keeps the same layout unless rewrite again\end{tabular} & \textless{}5\%~\cite{hawkins2017zipr} \\
\hline
SR~\cite{conti2016selfrando} & \begin{tabular}[c]{@{}l@{}}Function-level\\ randomization\end{tabular} & \begin{tabular}[c]{@{}l@{}}Load time\\reorder\end{tabular} & No & \begin{tabular}[c]{@{}l@{}}- Adds a linker wrapper that intercepts calls to the linker and asks the \\ \hspace{0.155cm}selfrando library to extract the necessary information to reorder functions\\ - Reorders functions every time when a binary is loaded into memory\end{tabular} & \textless{}1\%~\cite{conti2016selfrando} \\
%SR~\cite{conti2016selfrando} & \begin{tabular}[c]{@{}l@{}}Function-level\\ randomization\end{tabular} & \begin{tabular}[c]{@{}l@{}}Load time\\reorder\end{tabular} & No & \begin{tabular}[c]{@{}l@{}}- Adds a linker wrapper that intercepts calls to the linker and asks the \\ \hspace{0.155cm}selfrando library to extract the necessary information to reorder functions\\ - Reorders functions once a binary loaded into memory\\ - Reorders on each load\end{tabular} & \textless{}1\%~\cite{conti2016selfrando} \\
\hline
MCR~\cite{homescu2013profile} & \begin{tabular}[c]{@{}l@{}}Function- and register-\\level randomization\end{tabular} & \begin{tabular}[c]{@{}l@{}}Compile \& Link \\time reorder\end{tabular} & Yes & \begin{tabular}[c]{@{}l@{}}- Reorders functions and machine registers during link time optimization\\ - Implements compile-time randomization but defers compilation until \\\hspace{0.155cm}all translation units have been converted to bitcode\\ - Keeps the same layout unless compiled and built again\end{tabular} & 1\%~\cite{homescu2013profile} \\
\hline
CCR~\cite{koo2018compiler} & \begin{tabular}[c]{@{}l@{}}Function and block-\\level randomization\end{tabular} & \begin{tabular}[c]{@{}l@{}}Installation\\time\end{tabular} & Yes & \begin{tabular}[c]{@{}l@{}}- Extracts metadata during compilation\\ - Reorders functions and basic-block based on the metadata\\ - Keeps the same layout unless re-randomized again\end{tabular} & 0.28\%~\cite{koo2018compiler}
\end{tabular}
\end{table*}

%=======================================================================

%--------------------------------------------------------------------------------------
\subsection{Register corruption analysis} \label{appendix:register-corruption-analysis}
%--------------------------------------------------------------------------------------

Typically, a gadget contains a core instruction (other than {\em ret}) that serves the purpose of that gadget. For example, the core instruction of the gadget in Listing~\ref{payload-type1-gadgets} is {\em mov eax, edx} and the gadget serves as a move register (MR) gadget. The core instruction is the instruction that an attacker needs. All the instructions (except {\em ret}) before/after the core instruction is unnecessary. However, these extra instructions may modify the source/destination register of a core instruction. If these extra instructions modify the registers of a core instruction, we treat the gadget as a corrupted gadget. In Listing~\ref{payload-type1-gadgets}, the instruction ({\em mov edx, dword ptr [rdi]}) before the core instruction modifies the source register ({\em edx}) of the core instruction and the instructions ({\em shr eax, 0x10; xor eax, edx}) after the core instruction modify the destination register ({\em eax}). We identify three scenarios when core instructions get corrupted as follows:

\begin{enumerate}
    \item \textbf{\textit{Scenario 1}}: A core instruction is only affected by the instruction(s) before the core instruction,
    
    \item \textit{\textbf{Scenario 2}}: A core instruction is only affected by the instruction(s) after the core instruction, and
    
    \item \textbf{\textit{Scenario 3}}: A core instruction is affected by both the instruction(s) before/after the core instruction.
    
\end{enumerate}

\begin{figure}[!ht]
\centering
\begin{subfigure}{0.40\textwidth}
   \includegraphics[width=1\linewidth]{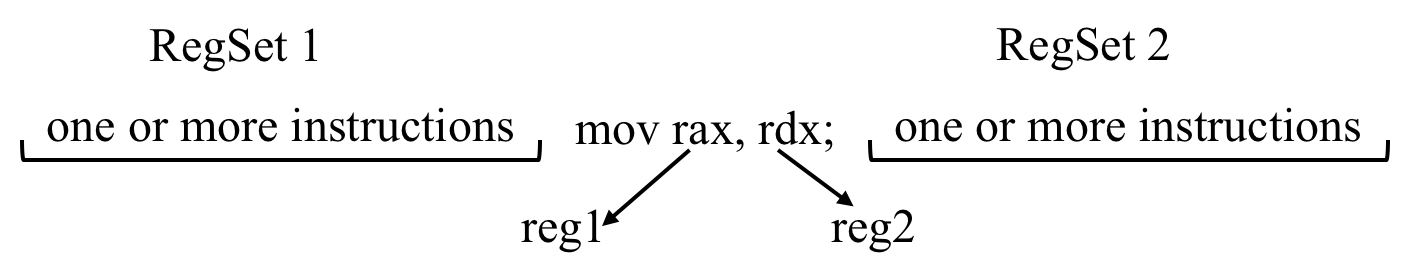}
   \caption{Type 1 gadget}
   \label{fig:payload_type1} 
\end{subfigure}

\begin{subfigure}{0.40\textwidth}
   \includegraphics[width=1\linewidth]{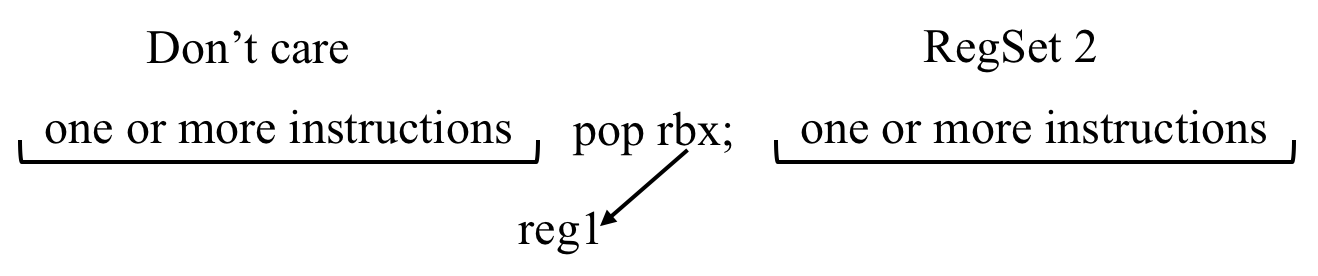}
   \caption{Type 2 gadget}
   \label{fig:payload_type2}
\end{subfigure}

\begin{subfigure}{0.5\textwidth}
   \includegraphics[width=1\linewidth]{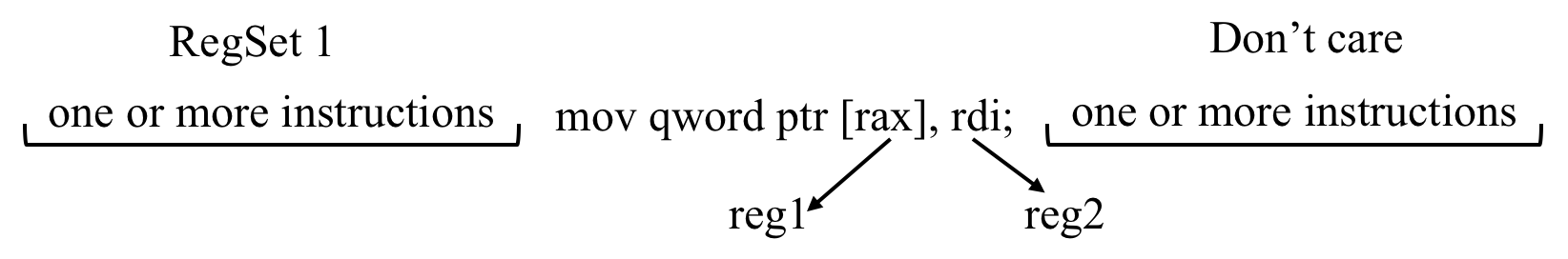}
   \caption{Type 3 gadget}
   \label{fig:payload_type3}
\end{subfigure}
\caption{A set of gadget types for measuring the quality of individual gadget through the register corruption analysis}
\label{fig-quality-gadget}
\end{figure}

For each gadget, we consider these three scenarios and determine whether the gadget is corrupted or not. We also identify three types of gadgets considering the three scenarios above where the core instruction can get corrupted. Figure~\ref{fig-quality-gadget} shows the three type of gadgets. Each gadget has one or more instructions before or after the core instruction. For example, Type 1 gadget in Figure~\ref{fig-quality-gadget} has a core instruction in the middle and one or more instructions before or after the core instruction. The core instruction has two registers for this kind. One or more instruction(s) before the core instruction may modify the source register ({\em rdx}) in Figure~\ref{fig:payload_type1}. Similarly, one or more instruction(s) after the core instruction may modify the destination register ({\em rax}) in the figure.

\lstset{style=mystyle}
\lstset{
    escapeinside={(*}{*)}
}
\lstset{language=[x64]Assembler}
\begin{lstlisting}[caption={An example gadget where the core instruction is ``mov eax, edx;".}, label={payload-type1-gadgets}]
mov edx, dword ptr [rdi]; (*\colorbox{light-gray}{\makebox(42,4){\textcolor{black}{mov eax, edx;}}}*) shr eax, 0x10; xor eax, edx; ret;
\end{lstlisting}

However, for Type 2 gadget in Figure~\ref{fig:payload_type2}, the core instruction has just one register. That means that the additional instructions before the core instruction cannot affect the register of the core instruction. Thus, we do not care the instructions before the core instruction. For Type 3 gadget in Figure~\ref{fig:payload_type3}, the core instruction writes the value of {\em rdi} to a memory location pointed by {\em rax}. That is why we do not care if the register ({\em rax}, {\em rdi}) values get modified by the instructions after the core instructions.

A gadget as corrupted if registers in the core instruction get modified. We perform our register corruption analysis by identifying the corrupted registers in the core instructions of a gadget as follows.

First, we identify the set of instructions (before/after the core instruction) that can modify the source/destination register of the core instruction. We find that 17 instructions ({\em mov}, {\em lea}, {\em add}, {\em sub}, {\em imul}, {\em idiv}, {\em pop}, {\em inc}, {\em dec}, {\em xchg}, {\em and}, {\em or}, {\em xor}, {\em not}, {\em neg}, {\em shl}, and {\em shr}) can modify a register value of a core instruction. That means that these instructions use the source register of a core instruction as its destination register or the destination register of a core instruction as its source register. We treat the registers of such instructions as conflicting registers.

%Second, we extract the conflicting registers for Types 1 and 3 gadgets. We call this set of registers as \texttt{RegSet1}. Similarly, we extract \texttt{RegSet2} for the instructions after a core instruction for Types 1 and 2.
Second, we extract the conflicting registers (\texttt{RegSet1}) for Types~1 and 3 gadgets and \texttt{RegSet2} for Types 1 and 2.
    
Third, if the \texttt{RegSet1} and/or \texttt{RegSet2} contain more than one conflicting registers, we treat the core instruction of that gadget as corrupted, i.e., the gadget itself is corrupted.
%\end{enumerate}

In this way, we measure the register corruption rate for \verb1MV1, \verb1LR1, \verb1AM1, \verb1LM1, \verb1AM-LD1, \verb1SM1, \verb1AM-ST1, \verb1SP1, and \verb1CALL1 gadgets.

%============ Reduction using a common set of apps and libs =================

\subsection{Validation of randomization results}\label{validation-same-set}
\begin{figure}[!ht]
  \centering
  \includegraphics[width=0.35\textwidth]{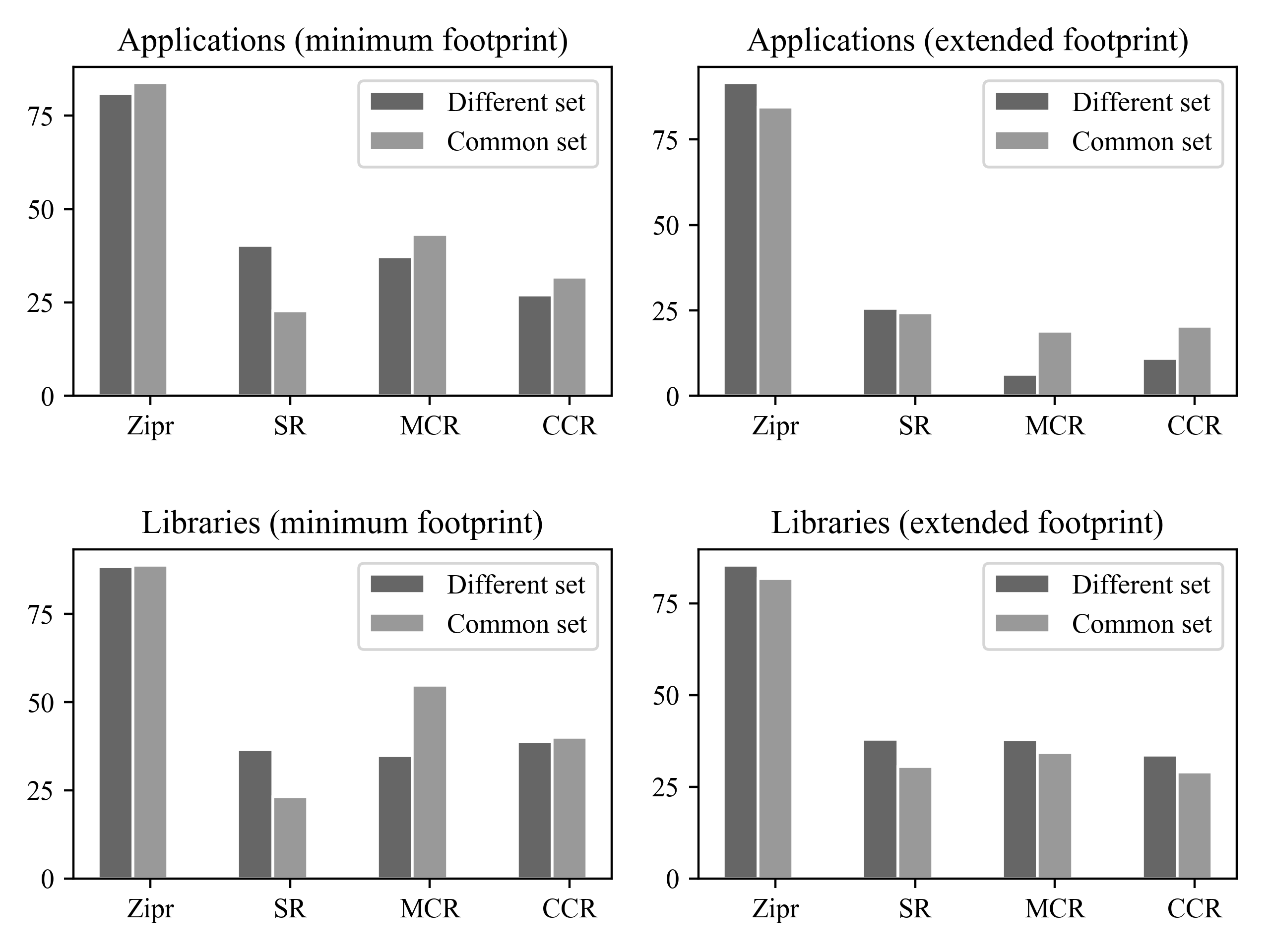}
  \caption{Reduction (\%) of TC gadgets observed for four (4) randomization tools using the common set of applications and libraries that the randomization tools can randomize.}
  \label{fig:reduction-same-set}
\end{figure}

We evaluate the randomization tools, i.e., Zipr~\cite{hawkins2017zipr}, SR~\cite{conti2016selfrando}, MCR~\cite{homescu2013profile}, and CCR~\cite{koo2018compiler} using the common set of applications and libraries that the four randomization tools can randomize. Figure~\ref{fig:reduction-same-set} shows the reduction of Turing-complete gadgets observed for four (4) randomization tools using the common set of applications and libraries. In most cases, the reduction using a different set of applications and libraries is similar to the reduction using a common set of applications.

%CCSR202 \subsection{Impact of leave Instruction}\label{leave-break-gadget}
%CCSR202 A {\em leave} instruction unfolds to {\em mov rsp, rbp; pop rbp}. So, the addition of a {\em leave}  instruction in a gadget modifies the stack pointer (SP), which affects the control flow of the resulting gadget chain. Whenever a {\em leave} is used in a gadget, it resets the stack pointer ({\em rsp}) by the value of the base pointer ({\em rbp}) using {\em mov rsp, rbp} instruction and restores the old {\em rbp} value from the stack using {\em pop rbp} instruction. That means that we need to control the value of {\em rbp} first, which requires efforts. Overall, using {\em leave}  instruction in a gadget requires additional effort and care.

\begin{figure}[!ht]
  \centering
  \includegraphics[width=0.35\textwidth]{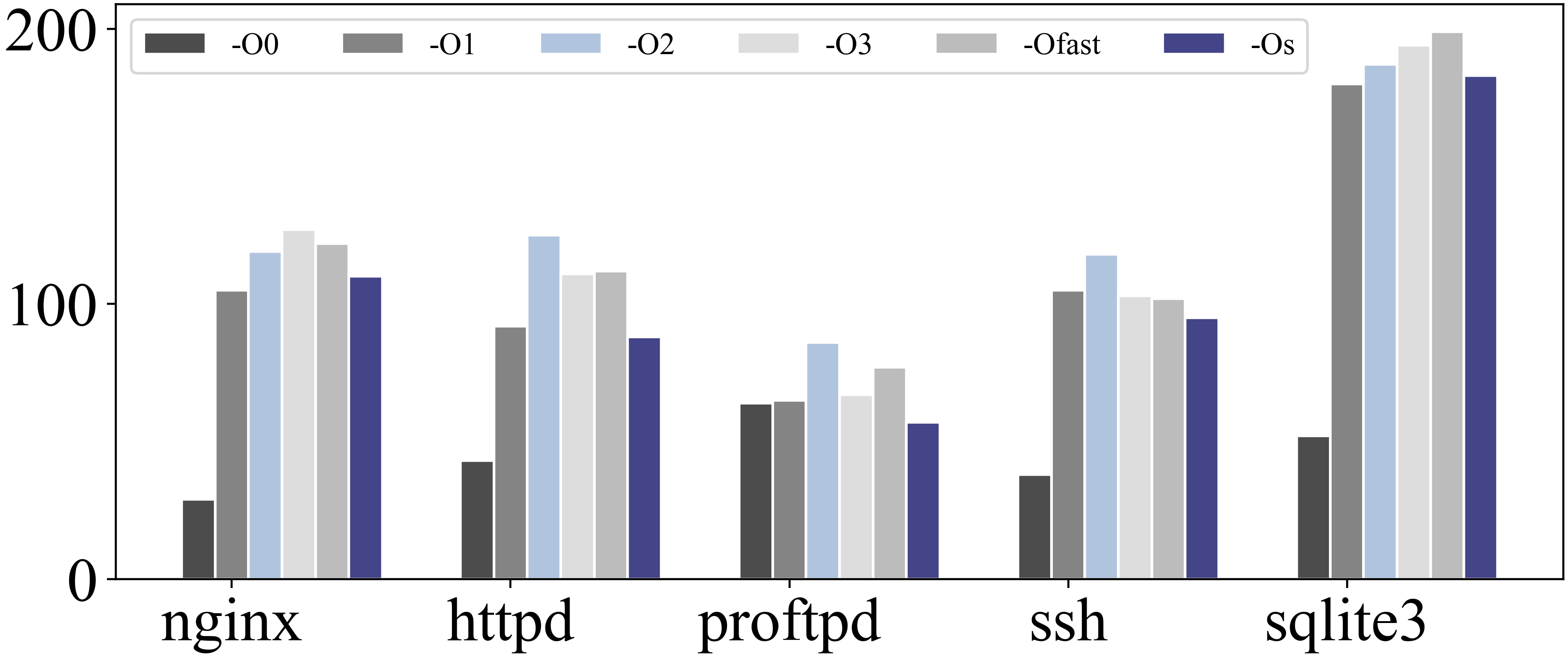}
  \caption{The number of Turing-complete minimum footprint gadgets at different optimization levels for GCC.}
  \label{fig-optim-plot}
\end{figure}

%==========================================

%===========ILR figure =============
\begin{figure}[!hb]
    \centering
    \includegraphics[width=0.45\textwidth]{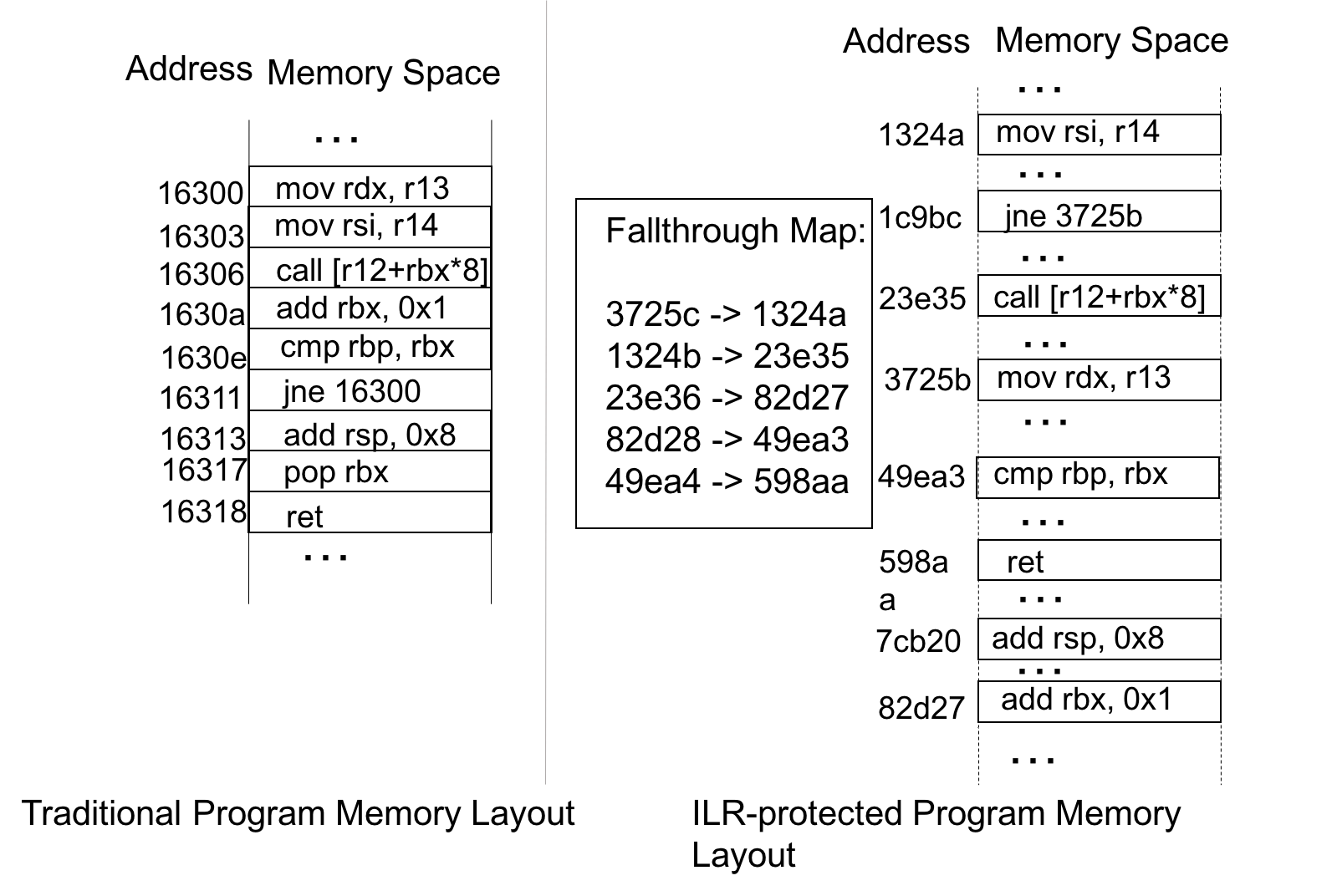}
    \caption{Instruction location randomization. This figure is adopted from ILR~\cite{hiser2012ilr}.}
    \label{fig:ilr-protected-program}
\end{figure}

%======================== Register corruption table =============================
\begin{table*}[!ht]
\centering
\scriptsize
\caption{Register corruption for various gadgets. The numbers before and after the vertical bar (\textbar) represent the average number of unique register usage and register corruption rate in a gadget, respectively. CG $\rightarrow$ Coarse-grained. FG $\rightarrow$ Fine-grained. Fine-grained versions prepared using SR~\cite{conti2016selfrando}.}
\label{register-corruption}
\renewcommand{\arraystretch}{1.1}
\setlength\tabcolsep{4.5pt} % default value: 6pt
\begin{tabular}{|l|l|l|l|l|l|l|l|l|l|l|c|l}
\cline{1-12}
\parbox[t]{2mm}{\multirow{6}{*}{\rotatebox[origin=c]{90}{\textbf{CG }}}}  & \textbf{Program} & \textbf{MV} & \textbf{LR} & \textbf{AM} & \textbf{LM} & \textbf{AM-LD} & \textbf{SM} & \textbf{AM-ST} & \textbf{SP} & \textbf{CALL} & Average & \\ \cline{2-12}
& Nginx & 4 \textbar\ 11\% & 2 \textbar\ 0.3\% & 3 \textbar\ 21\% & 3 \textbar\ {44}\% & 3 \textbar\ 6\% & 2 \textbar\ {47}\% & 2 \textbar\ 13\% & 2 \textbar\ 6\% & 2 \textbar\ 9\% & \textemdash &  \\ %\cline{2-11} 
& Apache & 4 \textbar\ 16\% & 2 \textbar\ 0.5\% & 3 \textbar\ 37\% & 2 \textbar\ 26\% & 3 \textbar\ 10\% & 2 \textbar\ 24\% & 2 \textbar\ 5\% & 2 \textbar\ 3\% & 2 \textbar\ 7\%  & \textemdash &  \\ %\cline{2-11}
& ProFTPD & 3 \textbar\ 69\% & 2 \textbar\ 0.6\% & 3 \textbar\ 7\% & 2 \textbar\ 24\% & 2 \textbar\ 20\% & 2 \textbar\ 16\% & 2 \textbar\ 11\% & 4 \textbar\ 1\% & 1 \textbar\ 6\%  & \textemdash & \\ %\cline{2-11}
& \textbf{Average} & \textbf{4} \textbar\ \textbf{32}\% & \textbf{2} \textbar\ \textbf{0.5}\% & \textbf{3} \textbar\ \textbf{21.7\%} & \textbf{2} \textbar\ \textbf{31.3}\% & \textbf{3} \textbar\ \textbf{12\%} & \textbf{2} \textbar\ \textbf{29\%} & \textbf{2} \textbar\ \textbf{9.7\%} & \textbf{3} \textbar\ \textbf{3.3\%} & \textbf{2} \textbar\ \textbf{7.3\%}  & 3 \textbar\ 16.3  &  \\ \cline{2-12} \cline{1-12}

\parbox[t]{2mm}{\multirow{5}{*}{\rotatebox[origin=c]{90}{\textbf{FG}}}} 
& Nginx & 3 \textbar\ 9\% & 1 \textbar\ 0.1\% & 2 \textbar\ 0.1\% & 3 \textbar\ {15}\% & 2 \textbar\ 45\% & 2 \textbar\ {13}\% & 2 \textbar\ 47\% & 1 \textbar\ 7\% & 2 \textbar\ 4\%  & \textemdash & \\ %\cline{2-11}
& Apache & 3 \textbar\ 27\% & 1 \textbar\ 1\% & 3 \textbar\ 41\% & 3 \textbar\ 27\% & 2 \textbar\ 19\% & 2 \textbar\ 41\% & 2 \textbar\ 0\% & 2 \textbar\ 2\% & 3 \textbar\ 27\%  & \textemdash & \\ %\cline{2-11}
& ProFTPD & 3 \textbar\ 14\% & 2 \textbar\ 1\% & 3 \textbar\ 4\% & 2 \textbar\ 19\% & 2 \textbar\ 22\% & 2 \textbar\ 35\% & 2 \textbar\ 6\% & 3 \textbar\ 11\% & 3 \textbar\ 28\%  & \textemdash & \\ %\cline{2-11}
& \textbf{Average} & \textbf{3} \textbar\ \textbf{16.7\%} & \textbf{1} \textbar\ \textbf{0.7\%} & \textbf{3} \textbar\ \textbf{15\%} & \textbf{3} \textbar\ \textbf{20.3\%} & \textbf{2} \textbar\ \textbf{28.7\%} & \textbf{2} \textbar\ \textbf{29.7\%} & \textbf{2} \textbar\ \textbf{17.7\%} & \textbf{2} \textbar\ \textbf{6.7\%} & \textbf{3} \textbar\ \textbf{19.7\%}  & 2 \textbar\ 17.24 & $\sim$5.7\%$\Uparrow$\\ \cline{1-12}%\hline
\end{tabular}
\end{table*}
%=============================================================================

%================== Tree of existing attacks and defenses =====================

\begin {figure*}[!b]
\centering
\begin{adjustbox}{width=0.95\textwidth}
\begin{tikzpicture}[sibling distance=12em, level distance=30pt,
  every node/.style = {shape=rectangle, rounded corners,
    draw, align=center,
    top color=white,font=\tiny},
    be/.style={circle,thick,draw,font=\tiny},
    emph/.style={edge from parent/.style={red,very thick,draw}},
    norm/.style={edge from parent/.style={black,thin,draw}}
    ]]
  \node (S) {CFI}
    child[red] { 
        node[black,draw] at (0,0.5) (A) {ASLR} 
        child {
            node[draw, black] (C) {Re-randomization (TASR~\cite{bigelow2015timely},\\ Shuffler~\cite{williams2016shuffler}, Remix~\cite{chen2016remix})}
            child [red] {
                node[draw,black] (E) [be][label={[black]below:$AC_1$: Vulnerable to \\simple ROP attacks}] {}
            }
            child[black] {
                node[draw] (F) at (-2,0) [be][label=below:$AC_2$: Vulnerable to simple ROP\\attacks if re-randomization time \\window is longer than the attack time] {}
            }
        }
        child {
            node[draw,black] (D) at (-0.25,0) {Memory protection + CPI or\\DPI (XnR~\cite{backes2014you}, NEAR~\cite{werner2016no},\\ Readactor~\cite{crane2015readactor}, Heisenbyte~\cite{tang2015heisenbyte},\\Oxymoron~\cite{backes2014oxymoron}, ASLR-Guard~\cite{lu2015aslr})}
            child[red] {
                node[draw,black](G) at (1.35, -0.45) {Re-randomization (TASR~\cite{bigelow2015timely},\\ Shuffler~\cite{williams2016shuffler}, Remix~\cite{chen2016remix})}
                child {
                    node[draw,black] (I) [be][label={[black]below:$AC_3$: Vulnerable to JIT-ROP~\cite{snow2013just}\\ and BROP~\cite{bittau2014hacking} type attacks}] {}
                }
                child[black] {
                    node[draw] (J) [be][label=below:$AC_4$: Vulnerable to JIT-ROP~\cite{snow2013just} type attacks if re-randomization\\time window is longer than the attack time] {}
                }
            }
            child[black] {
                node[draw] (H) [be][label=below:$AC_5$: Vulnerable to AOCR~\cite{rudd2017address} \\and CROP~\cite{gawlik2016enabling} type attacks] {}
            }
        }
    }
    child { node[draw] at (0.35, 0.6) (B) [be][label=below:$AC_6$: Prevents ROP-based\\attacks but vulnerable to\\data-only attacks~\cite{ispoglou2018block,hu2016data}] {}
      };

    \begin{scope}[nodes = {draw = none}]
    \path (S) -- (A) node [near start, left, yshift=-2pt]  {No};
    \path (S) -- (B) node [near start, right, yshift=-1pt]  {Yes};
    \path (A)     -- (C) node [near start, left, xshift=3pt]  {Coarse};
    \path (A)     -- (D) node [near start, right, xshift=-3pt]  {Fine};
    \path (C)     -- (E) node [near start, left, xshift=3pt]  {No};
    \path (C)     -- (F) node [near start, right, yshift=-3pt, xshift=-6pt]  {Yes};
    \path (D)     -- (G) node [near start, left, yshift=-2pt, xshift=5pt]  {No};
    \path (D)     -- (H) node [near start, right, yshift=-3pt]  {Yes};
    \path (G)     -- (I) node [near start, left, yshift=-2pt]  {No};
    \path (G)     -- (J) node [near start, right, yshift=-2pt]  {Yes};
  \end{scope}
\end{tikzpicture}
\end{adjustbox}
\caption{High-level view of the types of ROP attacks and attack-paths based on various security measures. Each rectangle and circle indicate security measures and attack types, respectively. AC stands for attack condition. All the attack conditions have W$\oplus$X, PIE, Canary, and RELRO implicitly.
}
\label{attac-decsision}
\end{figure*}
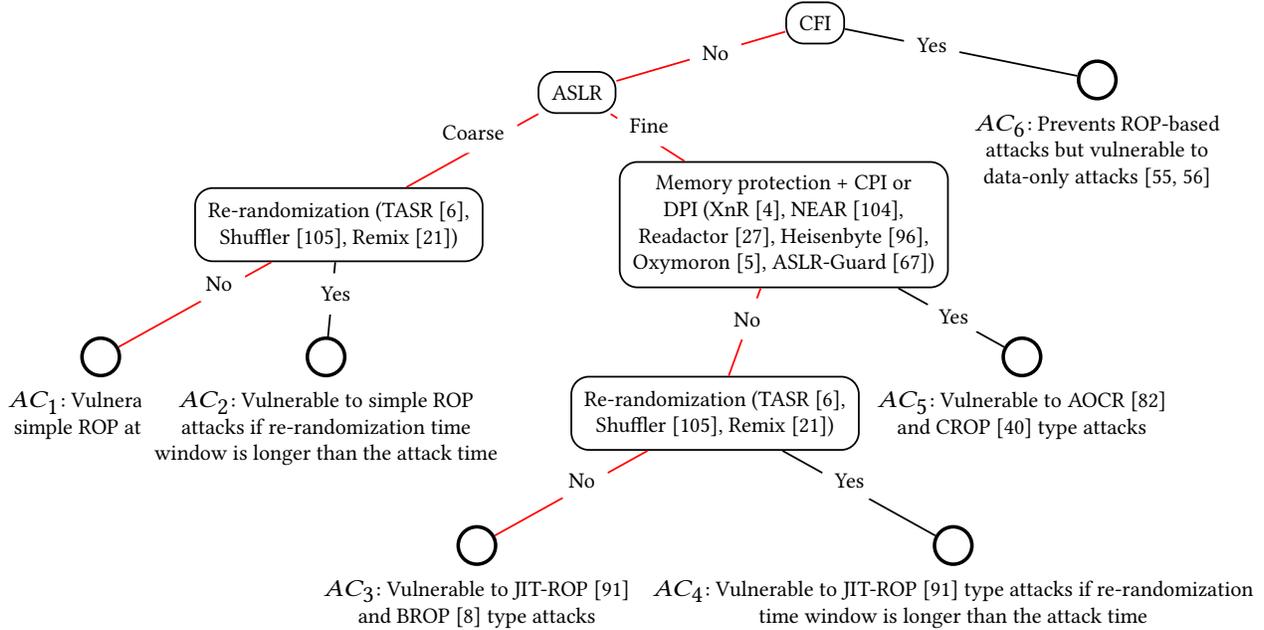

%====================================================

%\includepdf[pages=-,pagecommand={},width=\textwidth]{ccs-jan-2020/revision-for-CCS-2020.pdf}

\end{document}